\documentclass{aa}  

\usepackage{graphicx}
\usepackage{natbib,twoopt}
\usepackage{color}
\usepackage{lscape}
\usepackage{longtable}
\usepackage{times}
\usepackage{nccmath}
\usepackage{float}
\usepackage{xcolor}

\begin{document}

   \title{\textcolor{blue}{ Dust attenuation  and H$\alpha$ emission  in  a sample of    galaxies  observed with {\it Herschel} at $0.6 < z <1.6$}}

   \author{Buat V.
          \inst{1}
          \and Boquien, M.    \inst{2}
          \and Ma\l{}ek, K. \inst{1,3}
          \and Corre, D. \inst{1}
          \and Salas, H. \inst{2}
          \and Roehlly, Y. \inst{1,4,5}
          \and Shirley, R. \inst{5}
          \and Efstathiou, A. \inst{6}      
          }

   \institute{Aix Marseille Univ, CNRS, CNES, LAM Marseille,  France
\\
              \email{ veronique.buat@lam.fr}
             \and Centro de Astronom\'ia (CITEVA), Universidad de Antofagasta, Avenida Angamos 601, Antofagasta, Chile
              \and National Centre for Nuclear Research, ul. Hoza 69, 00-681 Warszawa, Poland
              \and Univ. Lyon1, ENS de Lyon, CNRS, Centre de Recherche Astrophysique de Lyon 
UMR5574, 69230, Saint-Genis-Laval, France
              \and Astronomy Centre, Department of Physics and Astronomy, University of Sussex, Falmer, Brighton BN1 9QH, UK
              \and School of Sciences, European University Cyprus, Diogenes Street, Engomi, 1516 Nicosia, Cyprus
             }



  \abstract
   {Dust attenuation shapes the spectral energy distribution of galaxies. It is particularly true for dusty galaxies in which stars experience a heavy attenuation. The combination of UV-to-IR photometry {with the spectroscopic measurement of the H$\alpha$ recombination line} helps to quantify  dust attenuation of the whole stellar population and its wavelength dependence.}
   {We want to derive the shape of the global attenuation curve and the amount of obscuration affecting young stars or nebular emission and the bulk of the stellar emission in a representative sample of galaxies selected in IR. We will compare our results to the commonly used recipes of Calzetti et al. and Charlot and Fall, and to predictions of radiative transfer models.  }
   {We selected an IR complete sample of galaxies in the COSMOS 3D-HST CANDELS field detected with the Herschel satellite with a signal to noise ratio larger than five. Optical to NIR photometry is available as well as NIR spectroscopy for each source. We reduced the sample to the redshift range $0.6 < z < 1.6$ to include the H$\alpha$ line in the G141 grism spectra. We have used a new version of the CIGALE code to fit simultaneously the continuum and H$\alpha$ line emission of the 34 selected galaxies.}
  {Using flexible attenuation laws with free parameters, we are able to measure the shape of the attenuation curve for each
galaxy as well as the amount of attenuation of each stellar population, the former being in general steeper than the starburst law in the
UV-optical with a large variation of the slope among galaxies. The attenuation of young stars or nebular continuum is found on average
about twice the attenuation affecting older stars, again with a large variation. Our model with power-laws, based on a modification
of the Charlot and Fall recipe, gives results in better agreement with the radiative transfer models than the global modification of the slope of the Calzetti law.}
   {}

   
   \maketitle
 %

\section{Introduction}

Modeling the spectral energy distribution (SED) of galaxies is a method  commonly used to derive physical parameters useful to  quantify  galaxy evolution, the most popular ones being the star formation rate (SFR) and the stellar mass ($\rm M_{star}$). The basics of these methods is to reconstruct the stellar emission of a galaxy with population synthesis models and  star formation histories of varying complexity with free parameters \citep[e.g.,][]{Walcher11, Conroy13}. Dust plays a crucial role  by strongly affecting and reshaping the spectral energy distribution (SED): it absorbs and scatters stellar photons, and thermally emits the absorbed energy in the infrared (IR) ($\lambda \sim 1-1000 ~\mu$m).  As a consequence any SED modeling must account for dust attenuation. The easiest way to do this  is to introduce a single attenuation curve which accounts for the complex blending of dust properties and relative geometrical distribution of stars and dust within a galaxy. The attenuation  law built by Calzetti and collaborators for nearby starburst galaxies \citep{Calzetti94, Calzetti00}  is by far the most commonly used. It is characterized by a greyer slope than both Milky Way  and Large Magellanic Cloud  extinction curves and by the  lack of the {so-called UV bump corresponding} to the 2175 \AA~absorption feature.  \citet{Charlot00} proposed a recipe also implemented in SED fitting codes. This recipe was originally built to be consistent with the properties of the nearby starburst galaxies analyzed  by \citet{Calzetti94} but both recipes differ  substantially in the visible-to-NIR  \citep[][Malek et al. 2018, in press]{Chevallard13, LoFaro17}.

The universality of an attenuation recipe  remains an open question in the nearby universe as well as for more distant galaxies.    \citet{Battisti16, Battisti17}  combined ultraviolet (UV) photometric data from GALEX with SDSS spectroscopy and near-IR (NIR) data from the UKIRT and 2MASS surveys  for several thousands of local galaxies, deriving an attenuation curve  similar to the starburst law although slightly lower in the UV. Conversely \citet{Salim18} extended the GALEX and SDSS data to the IR, with WISE and {\it Herschel}  detections and concluded that there is  a high variety of attenuation curves whose slope is a strong function of optical opacity confirming the results of radiative transfer calculations \citep[e.g.,][]{ Witt00, Pierini04, Tuffs04, Inoue05, Panuzzo07, Chevallard13, Seon16}. \\

At higher redshift \citet{Reddy15}  found  an average  attenuation curve of z$\sim$2 galaxies  similar to  the law of \citet{Calzetti00} in the UV. More recently \citet{Cullen18} also obtained an attenuation curve similar in shape to the starburst law  for z$\sim$3.5 star-forming galaxies. However \citet{Salmon16} presented  strong statistical evidence for the non universality of dust attenuation laws at $z\sim 2$. \citet{LoFaro17} derived  flat attenuation curves for ultra luminous IR galaxies (ULIRGs). 
Both observational and modeling analyses of the infrared excess (IRX) versus the UV spectral slope  also support the variability of the attenuation curve of star forming galaxies from  steep shapes for young, low-mass galaxies to greyer curves for IR bright, dusty galaxies \citep[e.g.,][]{Salmon16,  LoFaro17, Popping17,Narayanan18}.

When  stellar continuum  and nebular line  emissions   are considered together, both  must  be corrected for dust attenuation and a differential attenuation has been  introduced by \citet{Calzetti00} and \citet{Charlot00}. The relative attenuation affecting  distinct populations and their related emission remains an open issue as does the attenuation law to apply to  each component \citep[e.g.,][]{Garn10-2,Wild11,Chevallard13, Kashino13, Price14, Reddy15, LoFaro17}. The  difference found between the attenuation of  nebular lines on one side and  the continuum stellar population on the other side by \citet{Calzetti00} in  nearby starbursts  was confirmed by several studies  at higher redshifts \citep{Garn10-2,Price14} but other analyses found   a more similar attenuation for  young and old populations \citep{Pannella15, Puglisi16}. Large variations were measured among galaxies by \citet{Reddy15}, correlating with the sSFR  of the galaxies. 

A widespread  method to measure a differential attenuation is to assume the  attenuation law of  \citet{Calzetti00} and to compare different measures of SFR obtained with UV, IR and H$\alpha$ emission separately \citep{Garn10-2,Mancini11, Wuyts11, Puglisi16}. Indeed, the simultaneous fit of  combined spectroscopic and photometric data ensures a full consistency between the available data  but  is not trivial to perform. A major issue is the sampling difference between spectra (even of low or moderate resolution) and  photometric fluxes:  the number of spectroscopic data overwhelms the photometric ones. It makes it very difficult  to find any meaningful calculation of the likelihood, on which  Bayesian methods rely to build the probability density functions of  parameters \citep{Pacifici12}.
Methods using Monte Carlo codes  are now developed to  simultaneously fit full resolution spectroscopy and photometry  \citep{Chevallard16, Fossati18}. 
An alternative approach is to extract  some spectroscopic information like equivalent widths \citep{Pacifici15}, fluxes of emission lines, or  age-sensitive spectral indices \citep{Boselli16}, which are added to the input photometric data. This puts stronger constraints on  the  different stellar populations as well as the amount of dust attenuation while avoiding the oversampling of full spectra.

 In this work we  aim at  fully characterizing  dust attenuation  for   an IR complete sample of galaxies for which 3D-HST spectra cover the wavelength range of  the   H$\alpha$ line \citep{Momcheva16}.  We select galaxies in the COSMOS field  detected by {\it Herschel} \citep{Pilbratt10}, with either PACS \citep{Poglitsch10} or SPIRE \citep{Griffin10} instrument, at least in two bands with a signal to noise ratio (SNR) larger than 5. It ensures a good measure of the total IR emission, which is crucial to measure both  SFR and dust attenuation robustly. Photometric data from \citet{Laigle16} and spectroscopic data from the 3D-HST survey are then added. Ensuring an H$\alpha$ observation for all the sources limits the redshift range from 0.6 to 1.6. The photometric data and H$\alpha$ fluxes are simultanously fitted with an updated version of the CIGALE fitting code. This homogenous fit  is used to derive attenuation curves and differential attenuation characteristics. 
 
 The construction of the dataset is detailed in Section 2. We describe our SED fitting method and the various attenuation recipes adopted in this study in Section 3. The results of our analysis are presented  in Section 4 and discussed in Section 5. A summary of the main results  of this work is presented in Section 6.

\section{Data selection in the COSMOS field}
\begin{figure*}
\includegraphics[width=18cm]{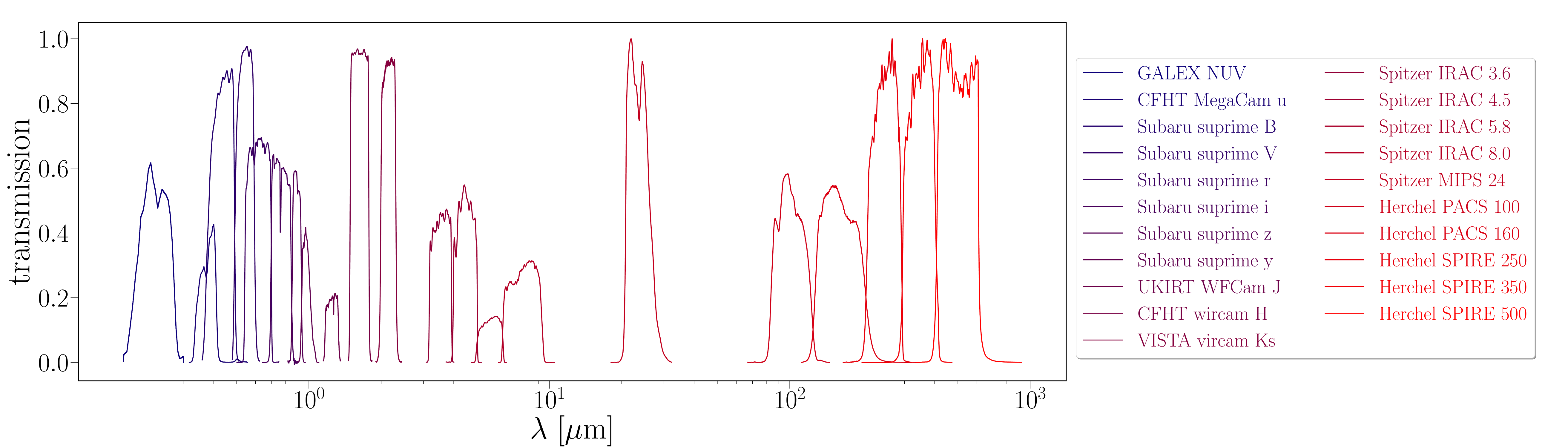}
\includegraphics[width=18cm]{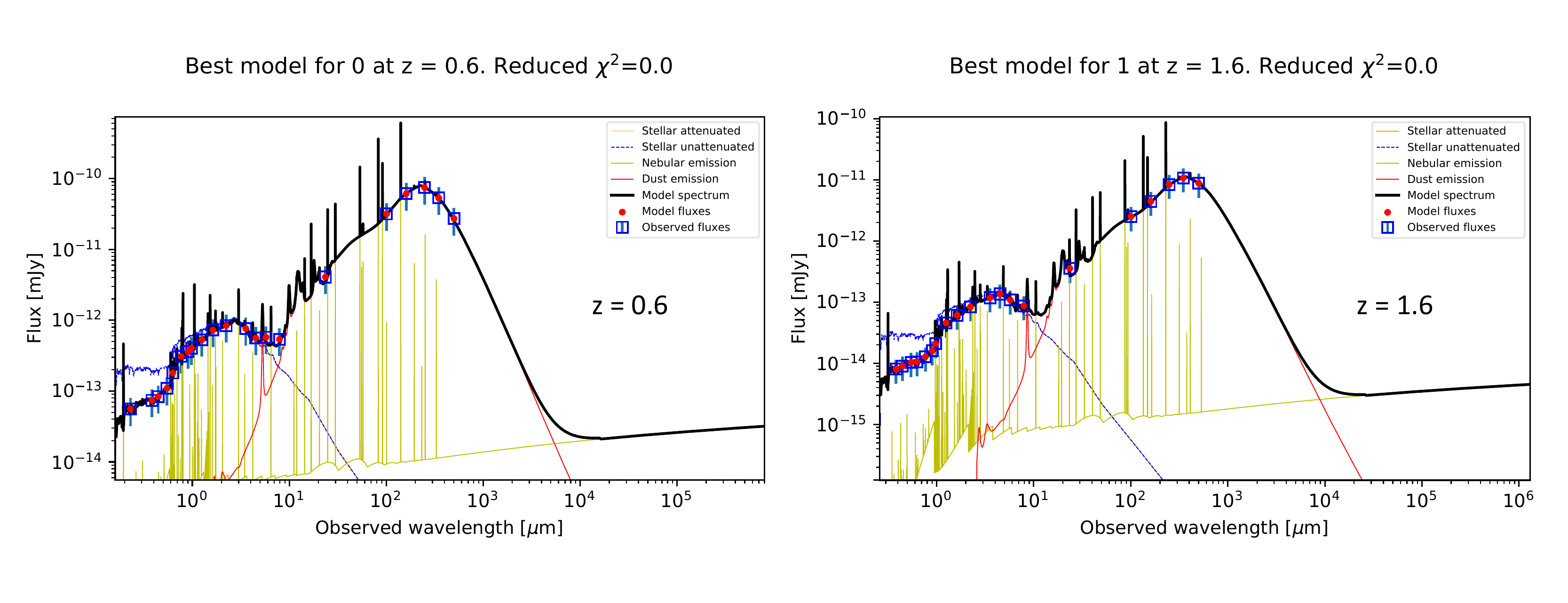}
\caption{{{\it Upper panel}: transmission curves of the filters used to build the SED of  galaxies, the list of the filters  is on the right side of the figure. {\it Lower panel}: template SED generated with CIGALE   at z=0.6 and z=1.6. The solid line is the SED computed with the same modules as the ones used for the fits, with a single model of star formation, dust attenuation and reemission. The red points are the values of the fluxes in the filters whose transmission curves are plotted in the upper panel, the blue squares are mock observations generated by the code with an error of 10$\%$ (blue vertical lines, at 3$\sigma$). The fluxes are calculated for a total star formation of one solar mass (standard mode for the generation of models with CIGALE).}}
	\label{filters}
\end{figure*}

\subsection{Target selection}

The HELP\footnote{The {\it Herschel} Extragalactic Legacy Project (HELP) is a European funded project to analyze all the cosmological fields observed with the Herschel satellite.}  collaboration provides homogeneous and calibrated multiwavelength data over  the {\it Herschel}  Multitiered Extragalactic Survey \citep[HerMES,][]{Oliver12}  and the  H-ATLAS survey (Eales et al. 2010) and some additional fields.  This unique dataset which covers   several hundred of square degrees is perfectly suited to build multiwavelength datasets with  well controlled selection functions. When combining data over a very large range of wavelengths, the source identification becomes a major issue due to the different spatial resolution and  source confusion. It is particularly true for {\it Herschel}  data with a beam reaching 36 arcsec at 500 $\mu$m. The  strategy adopted by HELP is to build a master list catalog of objects  as complete as possible for each field (Shirley et al., in preparation) and to use the NIR sources of this catalog as prior information for the {\it Herschel}  maps. The  XID+ tool \citep{Hurley17} uses a  Bayesian probabilistic framework  and works with prior positions. The code calculates the full posterior probability distribution on flux estimates. At the end a flux is  measured, in a probabilistic sense, for all the NIR sources of the master list.

In an early phase of the HELP project, the COSMOS field was chosen for a pilot study.  XID+ was run on {\it Spitzer} MIPS1,   {\it Herschel}  SPIRE   and PACS maps  with prior {\it Spitzer} IRAC positions for 694478 sources  with fluxes greater than 1 $\mu$Jy in any of the IRAC bands from the COSMOS2015 catalog of \citet{Laigle16}.
From this initial catalog we  select sources with at least either two PACS (at 100 and 160 $\mu$m) or  two   SPIRE  (at 250 and  350  $\mu$m) fluxes   with a SNR $>$ 5. The flux measurements  satisfy the criterion of goodness defined in XID+ and corresponding to a gaussian posterior distribution  of the estimated flux \citep{Hurley17}.  We are left with a sample of 2774 galaxies for the  two deg$^2$ of the COSMOS field.

We crossmatch our selected sources with the 3D-HST  catalog of the v4.1.5 release  with WFC3 G141 grism spectroscopy  over 122.2 arcmin$^2$ in the COSMOS field \citep{Momcheva16, Brammer12}.  We take  the best match within 2 arcsec from the  IRAC coordinates of our IR sources. 52 matches are found which correspond to our full {\it Herschel}  selection  within the area covered by the  HST grism survey (122.2 arcmin$^2$ against two  deg$^2$ covered by the COSMOS2015 catalog):  we get a complete IR selected sample with 3D-HST spectra for each source, all but one (resp two) sources are detected with PACS (resp SPIRE). We assign a redshift to each source with the $z_{best}$ value defined in the 3D-HST  (Cosmos-3dhst.v4.1.5.zfit.linematch) catalog. $z_{best}$ corresponds to a groundbased spectroscopic redshift for 11 sources and  to the best fit grism redshift for 39 sources. Two  galaxies have only a photometric redshift,  they are classified as AGN and will be excluded (see below). Since we are interested in H$\alpha$ line measurements we restrict the sample to the redshift range 0.6-1.6 where this line is observed with the G141 grism. 37 objects fulfill this condition. Then we exclude three  sources classified as AGN from the Chandra  Cosmos Legacy Survey\footnote{http://irsa.ipac.caltech.edu/data/COSMOS/tables/chandra/}. We are left with 34 galaxies, all but one galaxy, not detected with SPIRE, have PACS and SPIRE fluxes.

\subsection{Photometric data}
Starting with  the multiwavelength catalog of  \citet{Laigle16}, COSMOS2015, we add photometry from CFHTLS and  CFHT-WIRDS (Megacam bands u, g, r, i, z and WIRcam bands J, H, Ks), HSC-UDEEP and UDEEP (Hyper-Suprime-Cam bands g, r, i, z, y) and UKIDSS-LAS  (WFCAM bands J, H, K)  catalogs. We do this such that the final catalog  has one measurement per band per camera, that is   taking the order above we add measurements if and only if the object does not already have one in the given band from the given camera. The merging strategy is the same as described in Shirley et al. (in preparation) for the ELAIS-N1 field and is detailed in the 'dmu1\_ml\_COSMOS' product available on the Github repository of the project (https://github.com/H-E-L-P/dmu\_products/tree/master/dmu1/dmu1\_ml\_COSMOS). The best crossmatching radii are found to be 0.8 arcsec for CFHTLS, HSC-UDEEP and DEEP, and UKIDS, and 1 arcsec for CFHT-WIRDS. 

For the SED fitting with CIGALE we use only one measurement per band in order to avoid biasing the fit to parts of the spectrum with multiple cameras covering that band. We choose the ordering based on average depth on a given camera. This ordering, defining which camera measurement is preferred is: Megacam, Hyper-Suprime-Cam, VISTA, WIRCAM, WFCAM. For our 34 galaxies, all the photometric data come from the COSMOS2015 catalog. 
The photometric dataset consists of  16 bands from  GALEX NUV to {\it Spitzer} MIPS1 and of the 5 bands from {\it Herschel} PACS and SPIRE. With the H$\alpha$ fluxes we get 22 measurements per object. The NUV flux is measured  for only 24 sources, all the other bands are available for the whole sample, except for one galaxy  without any  SPIRE flux. The transmission curves of  the  filters  are presented in Fig. \ref{filters} {as well as the wavelength coverage on a  template SED for the two extreme redshifts of our sample (z=0.6 and 1.6).}

\subsection{Spectroscopic data}
We take the H$\alpha$  fluxes from the  3D-HST survey\footnote{https://3dhst.research.yale.edu/Data.php}.  \citet{vanDokkum11} showed that for a H$\alpha$  equivalent width larger than 10 \AA, the line is securely measured, without a significant contribution of the underlying stellar absorption. This absorption is directly taken into account in the measurement of the equivalent width \citep{Momcheva16} but in case of a faint emission line its contribution is large and increases the uncertainty on the flux measurement. The H$\alpha$ line  is detected with  SNR$<$2  for three of our sources,  and the observed equivalent width of two of these sources is lower than 10 \AA, all the others sources have  equivalent widths larger than 10 \AA.
The 3D-HST spectral resolution is not sufficient to detect separately the H$\alpha$ and the [NII] lines. For the sake of simplicity, hereafter we will refer as H$\alpha$ fluxes the sum of the H$\alpha$ and the two [NII] lines.

In this work we analyze individual spectral energy distributions in order to keep the diversity of the sources and  we do not  stack the data.  The H$\beta$ line is not detected on individual sources and we cannot measure Balmer decrements to directly estimate the dust attenuation affecting the Balmer lines \citep[e.g.,][]{Garn10,Price14}. The dust emission of our galaxies is measured thanks to  the {\it Herschel}  data and puts a strong constraint on the stellar obscuration through the energy budget between  stellar and dust emissions as explained in the next section.

\section{The SED fitting method}
\subsection{The code CIGALE}

The SED fitting is performed with the CIGALE code\footnote{https://cigale.lam.fr}. We refer to Boquien et al. (2018, submitted) and to the description on the web site of CIGALE for more details about the code. CIGALE combines a UV-to-NIR stellar SED with a dust component emitting in the IR and conserves the energy balance  between dust absorbed emission and its thermal reemission.  The nebular emission is added  from the  Lyman continuum photons produced by  the stellar component. The nebular continuum and the emission lines are calculated from a grid of nebular templates \citep{Inoue11} generated with CLOUDY 08.0. The intensity of 124 lines are included, the models are parametrized with the ionization parameter and the metallicity. Star formation histories as well as dust attenuation recipes can be taken  either free or fixed. 
The global quality of the fit is assessed by a reduced $\chi^2$ ( $\chi_r^2$) defined as the   $\chi^2$ divided by the number of input fluxes. This definition does not account for the degree of freedom of the fit, this quantity is difficult to determine  since the free parameters are not independent and the equations linking them are non linear \citep{Chevallard16}. The  $\chi_r^2$  value is only used as an indication of the global quality of the fit and the best model is not used for the estimation of the parameters. Instead, the values of the parameters and the corresponding uncertainties are estimated by taking the likelihood weighted mean and standard deviation, the probability distribution function (PDF)  of each parameter are built. Here, we only describe the assumptions and choices specific to our current study.

For the purpose of this work,  CIGALE  is modified  (version 0.12.1) so that  photometric and emission line fluxes (here only the H$\alpha$ line) fluxes  are fitted simultaneously.  \\
We assume a delayed star formation history (SFH) with the functional form $SFR \propto t \exp(-t/\tau$). A recent constant burst of star formation is overimposed whose amplitude is measured by the stellar mass fraction produced during the burst. This functional SFH is found to give satisfactory results for the galaxies of the ELAIS-N1 field  and is routinely used to fit all the HELP samples (Malek et al. 2018, in press). We adopt a Salpeter Initial Mass Function \citep{Salpeter55} and the stellar models of \citet{bc03}. The metallicity is fixed to the solar value. The input values characterizing the SFH are presented in Table \ref{param}.

\subsection{Parametric attenuation laws}

The major aim of this work is to analyze the dust attenuation and its variability among galaxies. Different attenuation recipes are implemented in CIGALE and we start here  with  two of the most popular ones.

The first recipe we consider  is based on  the  attenuation law for the stellar continuum of \citet{Calzetti00}. The nebular component  (and therefore the H$\alpha$ line) is extinguished with a simple screen model, a Milky Way extinction curve \citep{Cardelli89} and a color excess $E({\rm B-V})_{\rm line}$. The input parameters used  to quantify the amount of attenuation are  the color excess to apply to the nebular emission, $E({\rm B-V})_{\rm line}$, and the   ratio   $E({\rm B-V})_{\rm star}/E(B-V)_{\rm line}$, $E({\rm B-V})_{\rm star}$ is the color excess to apply to the whole stellar continuum. This ratio was found  equal to 0.44  for local starbursts  \citep{Calzetti01}. As we discussed in the introduction, the variability of this ratio is   questioned and  this parameter is taken free in our fitting procedure. This attenuation recipe (attenuation law of \citet{Calzetti00}, Milky Way extinction curve for the nebular component, and a variable ratio of color excesses) will be called C00 hereafter.

The  second recipe we consider is similar to the one proposed  by \citet{Charlot00}. It differs from C00 in its philosophy. A differential attenuation between young (age $<10^7$ years) and old (age $>10^7$ years) stars is assumed. Both young and old stars undergo  an attenuation in the interstellar medium (ISM), the young stars  are affected by an extra attenuation in the birth clouds (BC). Both  attenuation laws are modeled by a power law and normalized to the amount of attenuation in  the V band, $A^{\rm ISM}_{\rm V}$ and $A^{\rm BC}_{\rm V}$,
\begin{flalign}
\label{taubc_cf00}
& A^{\rm BC}_{\lambda} = A^{\rm BC}_{\rm V} (\lambda/0.55 )^{n^{\rm BC}}\\
& A^{\rm ISM}_{\lambda} = A^{\rm ISM}_{\rm V} (\lambda/0.55 )^{n^{\rm ISM}}.
\label{cf00}
\end{flalign}
\citet{Charlot00} fixed both exponents of the power laws $n^{\rm BC}$ and $n{\rm ISM}$  to -0.7, although a value of -1.3 is  initially introduced in their model and further adopted by \citet{daCunha08}. 
 $\mu$ is defined as the ratio  of  the attenuation in the V band  experimented by old and young stars:
 \begin{flalign}
 \mu = A^{\rm ISM}_{\rm V} /( A^{\rm ISM}_{\rm V} +A^{\rm BC}_{\rm V}).
 \end{flalign}
 \citet{Charlot00}  obtained $\mu = 0.3$   from their study of nearby starburst galaxies.  As for the ratio of the color excess in the C00 recipe, we take  $ \mu$  as  a free parameter in our fits. The recipe defined with the the exponents of the power laws $n^{\rm BC}$ and $n^{\rm ISM}$  equal to  -0.7 and $\mu$ taken free will be refered as CF00 hereafter.

We further introduce some flexibility in the previous recipes and allow the general shape of the attenuation law to be steeper or flatter than the original ones.  CF00  becomes the Double Power Law with free slopes \citep[hereafter DBPL-free, as introduced by][]{LoFaro17} :  $n^{\rm BC}$ and $n^{\rm ISM}$ are considered as free parameters. \\
 C00  is also modified and we define the  Calzetti-like recipe \citep{noll09}:
\begin{equation}
A(\lambda) = E({\rm B-V})_{\rm star} k^{\prime}(\lambda) \left(\frac{\lambda}{\lambda_{\rm V}}\right)^{\delta},
\label{dust1}
\end{equation}
$k^{\prime}(\lambda)$ comes from C00, $\delta$ is  a free parameter. \\
In Eq. \ref{dust1}, $\delta= 0$ corresponds to the original recipe C00 and  $E(B-V)_{\rm star}$ is equal to $A_{\rm B}-A_{\rm V}$. When  $\delta  \neq 0$,  $E(B-V)_{\rm star}$ from Eq.\ref{dust1} is no longer equal  to $A_{\rm B}-A_{\rm V}$. To avoid a wrong definition, the Calzetti-like recipe  is implemented  in CIGALE in order that the input parameter  $E(B-V)_{\rm star}$ is always equal to $A_{\rm B}-A_{\rm V}$, for any value of $\delta$ (Boquien et al. 2018, submitted).  We adopt a fine sampling of the input parameters related to dust  attenuation (Table \ref{param}) in order to get reliable PDFs, mean values and standard deviations.
\subsection{Fitting  the SEDs}

\begin{figure}
\includegraphics[width=8cm] {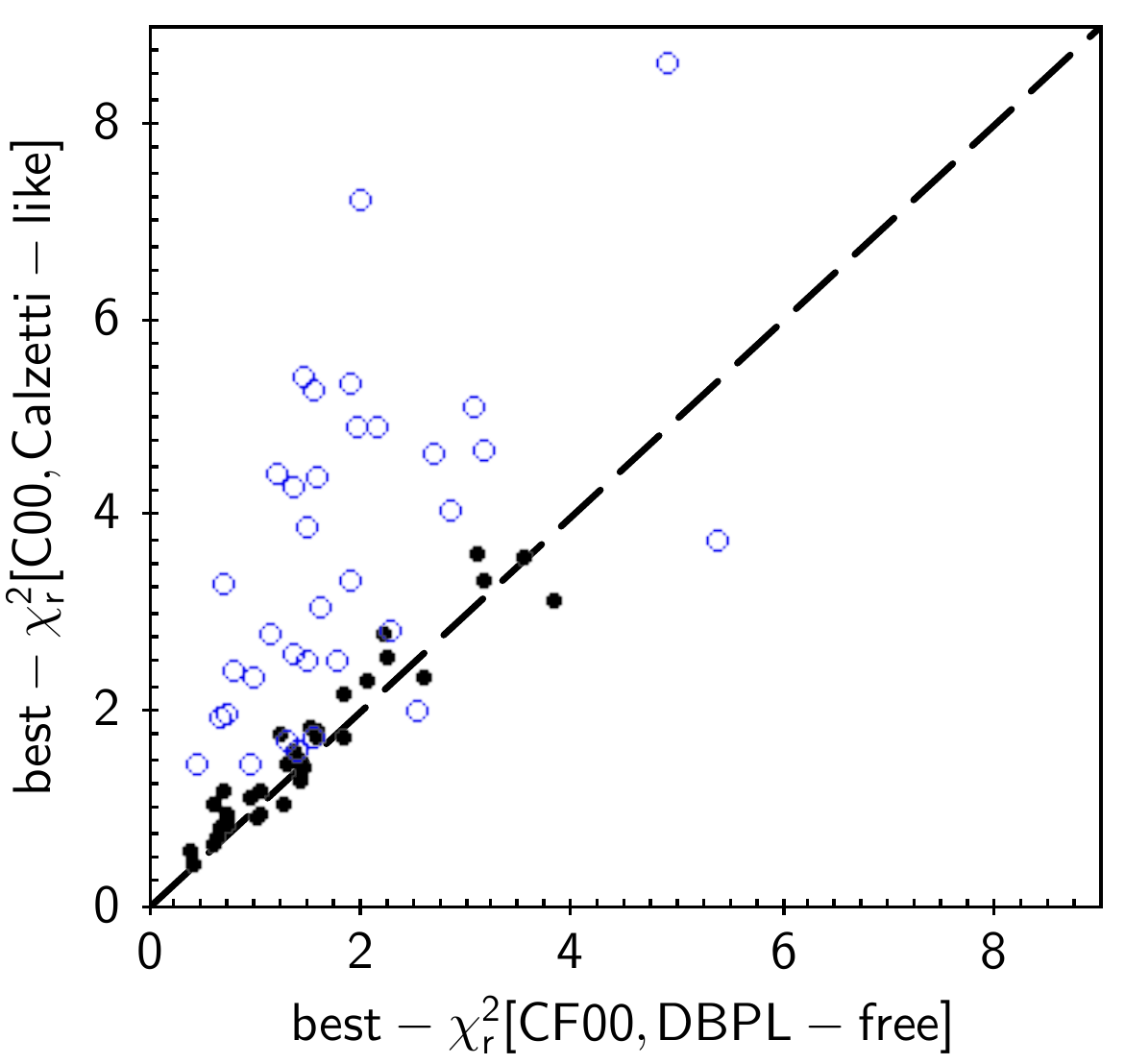}
\caption{Comparison of the quality of the fits with the four attenuation recipes considered.   The   best  $\chi_r^2$ obtained with the original  dust attenuation recipes CF00 (x axis) and C00 (y axis) are plotted with {blue} empty circles. The best  $\chi_r^2$ obtained with the flexible recipes  DBPL-free (x axis) and Calzetti-like (y axis) with the slopes of the attenuation laws taken as free parameters  are plotted with { black} filled circles.}
\label{chi2}
\end{figure} 

\begin{table*}
\scriptsize
\centering
\begin{tabular}{c c c}
\hline\hline
Parameter & Symbol & Range \\
\hline
\multicolumn{3}{c}{Delayed Star Formation History and  Recent Burst}\\
\hline
age  of the main population& $age_{main}$ & 2000,  4000, 6000, 7500\,Myr\\
$e$-folding timescale of the delayed  SFH & $\tau$ & 1000,3000,5000\,Myr\\
age of the burst &  $age_{burst}$ & 10, 20, 50, 70\,Myr\\
burst stellar mass fraction  & $f_{burst}$& 0.0, 0.001, 0.01\\
\hline
\multicolumn{3}{c}{Dust attenuation}\\
\hline
\textit{ C00 and Calzetti-like recipe:} & &\\
color excess of  nebular emission & $E(B-V)_{\rm line}$ & 0.05,0.1,0.15,0.2,0.25 and 0.3 to 1.9 mag per bin of 0.1 mag\\
color excess ratio between continuum and nebular emission& $E(B-V)_{\rm star}/E(B-V)_{\rm line}$ & 0.1 to 1  per bin of 0.05\\
slope of the power law modifying the attenuation curve & $\delta$  &{\bf 0}, -0.6 to 0.2 per bin of 0.1\\
\hline
\textit{CF00 and DBPL-free recipe:} & &\\
V-band attenuation in the ISM  & $A_{\rm V}^{\rm ISM}$ & 0.5 to 2.6 mag (per bin of 0.1 mag) \\
power law slope of dust attenuation in the BCs & $n{\rm BC}$ & -0.7\\
$A_{\rm V}^{\rm ISM}/ (A_{\rm V}^{\rm ISM}+A_{\rm V}^{\rm BC}$) & $ \mu$ & 0 to 1 (per bin of 0.1)\\
power law slope of dust attenuation in the ISM & $n^{\rm ISM}$ &  {\bf -0.7}, -1.2 to -0.4 (per bin of 0.1)\\
\hline
\multicolumn{3}{c}{Dust emission}\\
\hline
mass fraction of PAH & $q_{\rm PAH}$ & 1.12, 2.50, 3.19\\
minimum radiation field & U$_{\rm min}$ & 5., 10., 25.0\\
powerlaw slope dU/dM $\propto$ U$^{\alpha}$ & $\alpha$ & 2.0\\
dust fraction in PDRs & $\gamma$  & 0.02\\
\hline\hline
\end{tabular}
\caption{CIGALE modules and input parameters used for all the fits. The input values used for the CF00 and C00 recipes are in boldface type. The initial mass function is that of \citet{Salpeter55} }
\label{param}
\end{table*}

\begin{figure}
\includegraphics[width=\columnwidth] {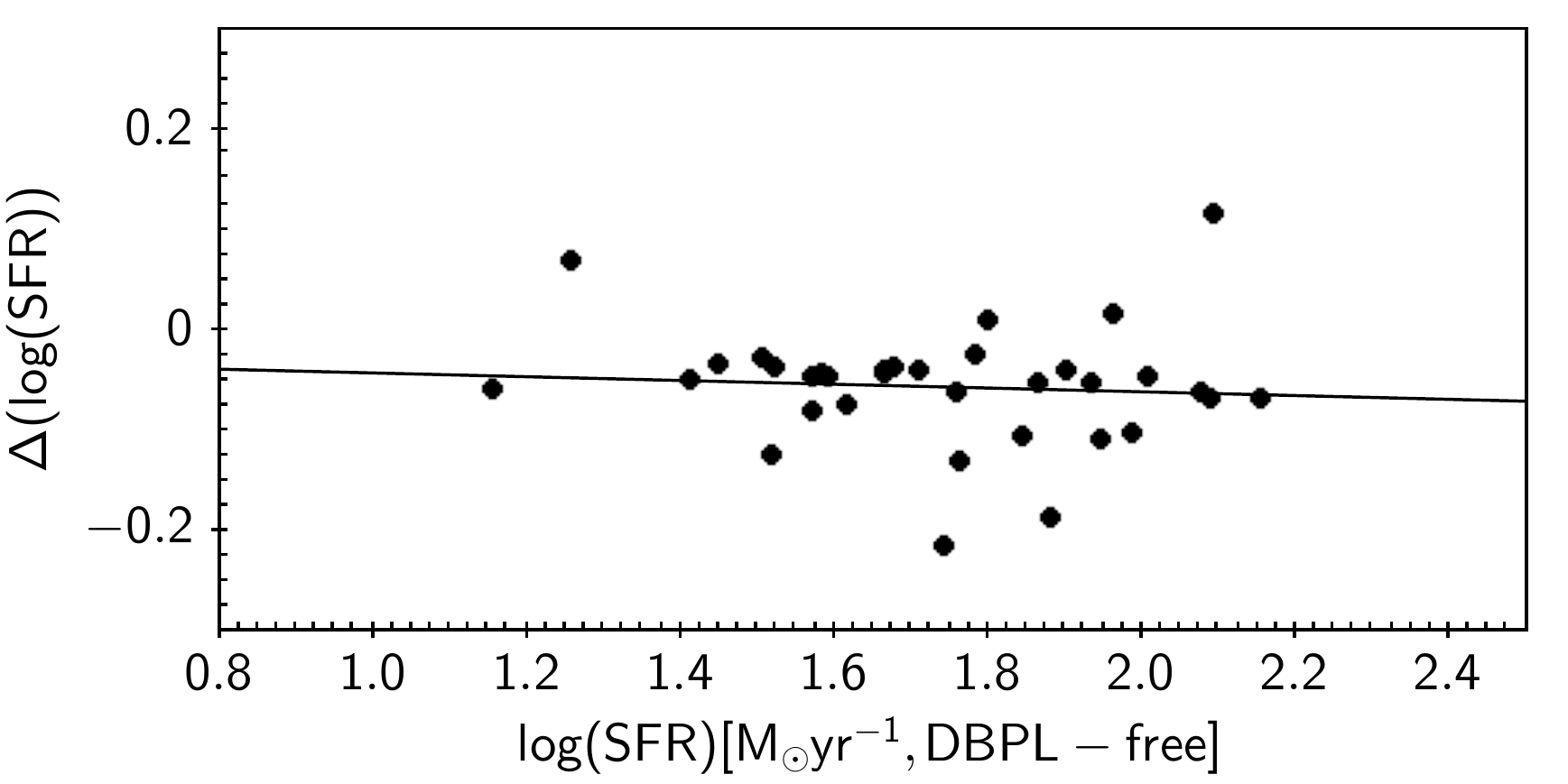}
\includegraphics[width=\columnwidth] {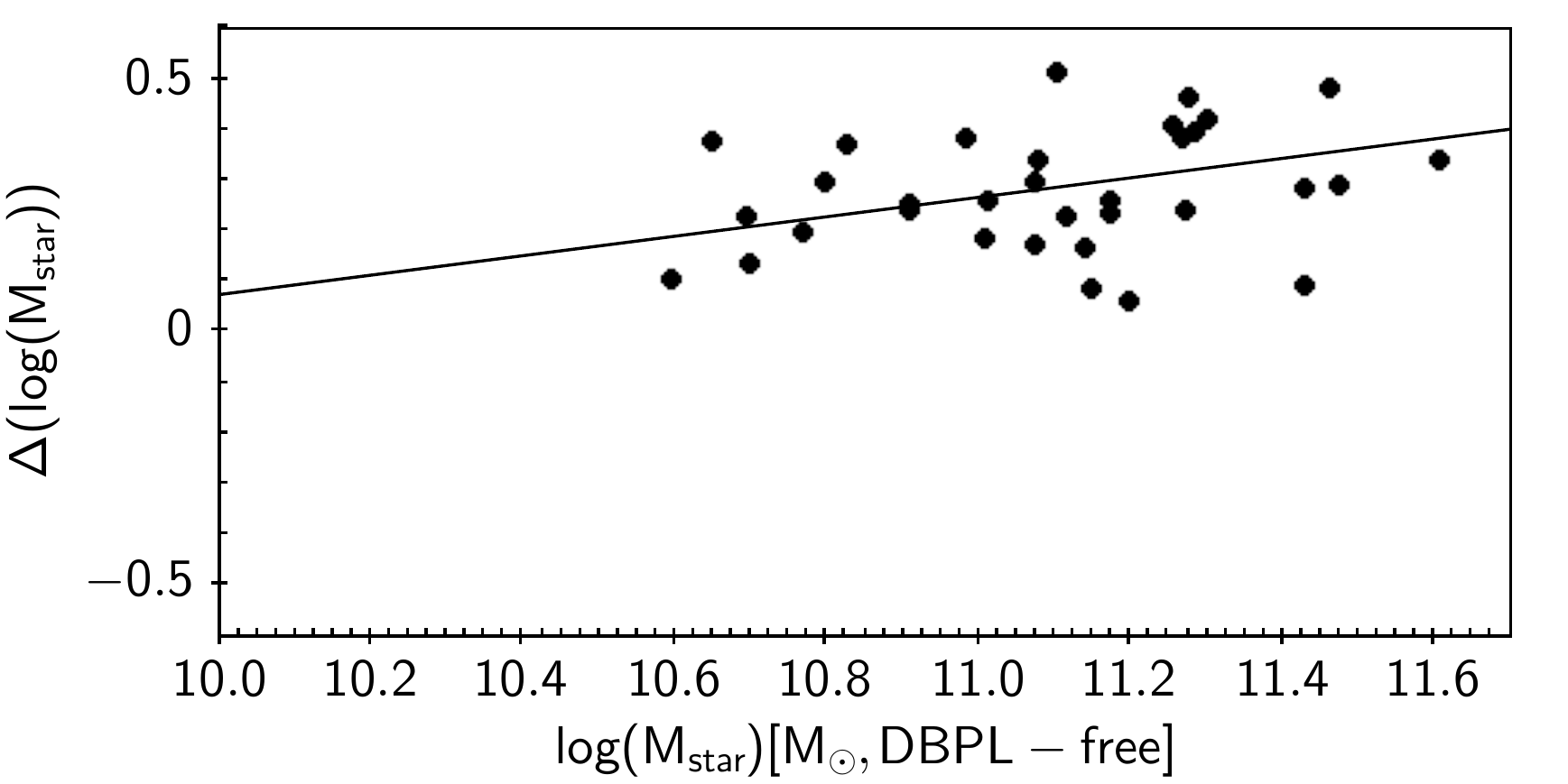}
\caption{ Comparison of SFR  and M$_{\rm star}$ for the two flexible recipes. The values along the x axis are obtained with the DBPL-free model, the difference between the DBPL-free and Calzetti-like models is plotted on the y axis (DBPL-free minus Calzetti-like logarithmic values). The solid lines are the result of a linear regression.}
\label{delta}
\end{figure} 

\begin{figure}
\includegraphics[width=\columnwidth] {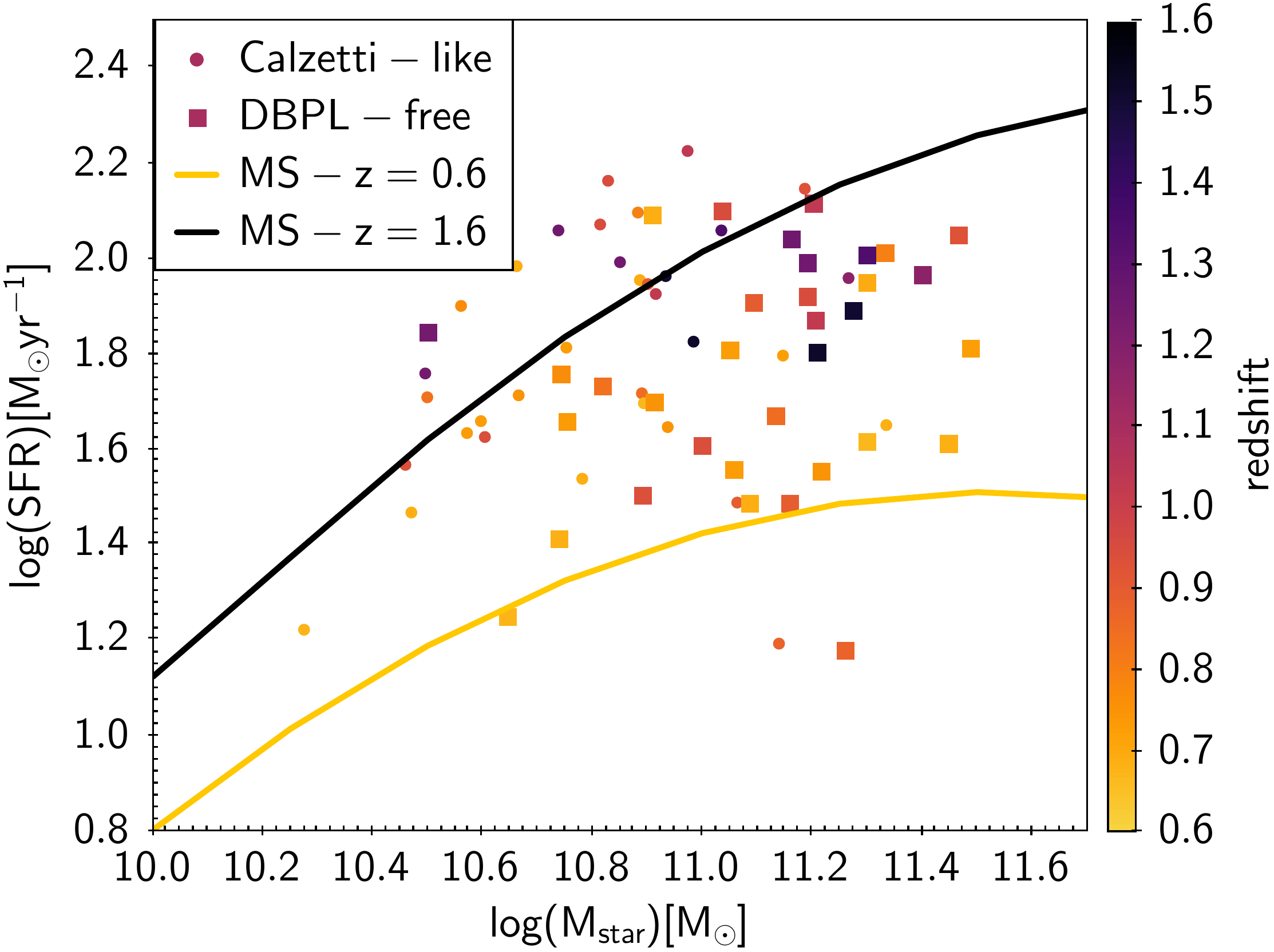}
\caption{SFR plotted against M$_{\rm star}$ for the two flexible recipes DBPL-free (filled squares) and Calzetti-like (dots). The relations of \citet{Schreiber15}  for redshift z=0.6 and z=1.6 are represented with solid lines, the redshift of the sources is color coded}
\label{MS}
\end{figure}

\begin{figure}
\includegraphics[width=\columnwidth] {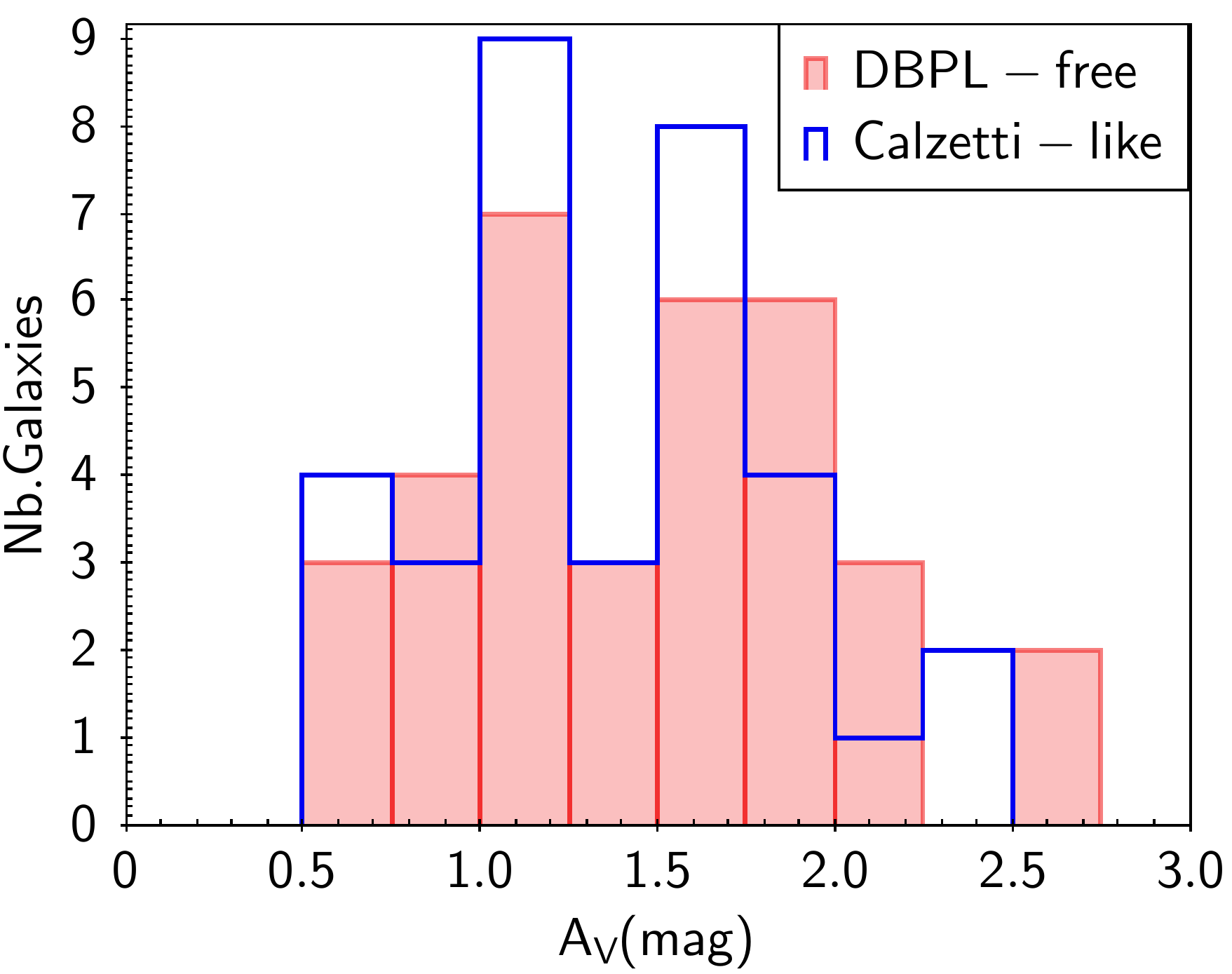}
\caption{Distribution of the total attenuation in the V band, for DBPL-free (filled red histogram) and Calzetti-like (blue empty histogram).}
\label{AVdist}
\end{figure} 

 The modules used for the fits and all  their  input values are summarized in Table \ref{param}. The dust  component is calculated  with the \citet{Draine07} models. We use the standard set of parameters adopted for our previous studies of galaxies  at similar redshift and selected in IR \citep[][Malek et al., 2018, in press]{LoFaro17}. In this work we will not  focus on the detailed IR emission, we are only interested in the measure of the total IR emission  to perform the energy budget of the  stellar photons  absorbed  and reemitted by dust. 
 
The values of  $\chi_r^2$  for the different attenuation recipes (CF00, C00, DBPL-free and Calzetti-like) are plotted in Fig. \ref{chi2}. The  global quality of the fits is  good, in particular with the flexible recipes DBPL-free and Calzetti-like, we will go back to the comparison of the models in the next section.

The SFR and dust luminosities   are found to be similar with all the  attenuation recipes with median differences lower than 0.05 dex {as illustrated in Fig. \ref{delta} for the SFR values obtained with the flexible recipes}. The dust luminosity spans only one decade  ($\rm 10^{11}-10^{12}L_{\sun}$). The mass fraction produced during the current burst is found very low, at most equal to $1\%$: a single delayed star formation describes the SFH of our galaxies quite well.
The stellar masses (M$_{\rm star}$) exhibit significant differences between models as already underlined by \citet{LoFaro17}:  M$_{\rm star}$ is  found higher with    CF00 and DBPL-free recipes because of a flatter attenuation curve in the visible-to-NIR  as it will be shown in section 4. {The  M$_{\rm star}$ values obtained with the two flexible recipes are compared in Fig. \ref{delta}, the median difference reaches  0.27 dex between DBPL-free and Calzetti-like recipes (0.22 dex between CF00 and C00 recipes).} When fixed and flexible recipes are compared,  the median difference is much lower (0.13 dex between C00 and Calzetti-like, 0.05 dex between CF00 and DBPL-free). The SFR and M$_{\rm star}$ values for our galaxies  obtained with the Calzetti-like and DBPL-free recipes are plotted in Fig. \ref{MS} together with  the average  relations of \citet{Schreiber15} for Main Sequence (MS) galaxies in the same redshift range.  The shift towards lower masses of the values obtained with the Calzetti-like method implies that the galaxies exhibit larger  {specific SFR (SFR divided by stellar mass, hereafter sSFR). At a given stellar mass no galaxy of our sample is found above the MS by  more than a  factor of 4, which corresponds to the definition of a starburst according to  \cite[e.g.,][]{Rodighiero11,Sargent12}. We conclude that none of our galaxies is starbursting, in agreement with the delayed SFH.}

When the DBPL-free recipe is used, it is not possible to estimate securely $n^{\rm BC}$  as already shown by \citet{LoFaro17} and we fix its value to -0.7,  we checked that using -1.3 does not modify any of the results of this study.  
 The distribution of the total attenuation in the V band is plotted in Fig. \ref{AVdist} for the two flexible recipes, similar distributions are found with the recipes CF00 and C00.

\section{The  derived attenuation laws}
\subsection{Testing  the original recipes of \citet{Charlot00} and \citet{Calzetti00}}

We first compare the results of the runs with the original recipes  C00 and CF00  (Fig. \ref{chi2}). The fits are significantly  better with CF00 with a median value of $\chi_r^2$ equal to 1.57 against 3.32 for C00.

The fits are  improved when the slope of the attenuation law is taken free and both recipes  return fits of similar quality with a  median  $\chi_r^2$ of 1.35 and 1.45 
 for DBPL-free and Calzetti-like recipes respectively. An improvement of the fits is expected since an additional free parameter is introduced in each recipe. However  the reduction of  $\chi_r^2$ is stronger  in the case of the Calzetti recipe, the CF00 modeling giving satisfactory results already in its original form. 
 {This is confirmed by the comparison of the values of the Bayesian Information Criterion (BIC) which accounts for the increase of free parameters\footnote{We apply the definition adopted by  \citet{Ciesla18}: ${\rm BIC}= \chi^2+k \times \ln(n)$. $\chi^2$ is the non-reduced of the best fit, $k$ the number of free parameters and $n$ the number of data fitted.}. The difference between  BIC$_{\rm C00}$ and BIC$_{\rm Calzetti-like}$ is found higher than 6 (which corresponds to a strong evidence against the model with the higher BIC) for $85\%$ of the sample, it drops to 44\% between  BIC$_{\rm CF00}$ and BIC$_{\rm DBPL-free}$.}
 It {implies } that the global shape of the attenuation curves resulting from the CF00 recipe is  more adapted to our galaxy sample  than the C00 attenuation law as also found by \citet{LoFaro17} and Malek et al. (2018, in press) on different HELP selected samples.
In the following we continue the analysis using  the flexible attenuation recipes.

\subsection{ The  attenuation laws derived with flexible recipes}
\begin{figure*}[!]
   \centering
   \includegraphics[width=0.3\hsize, angle=0]{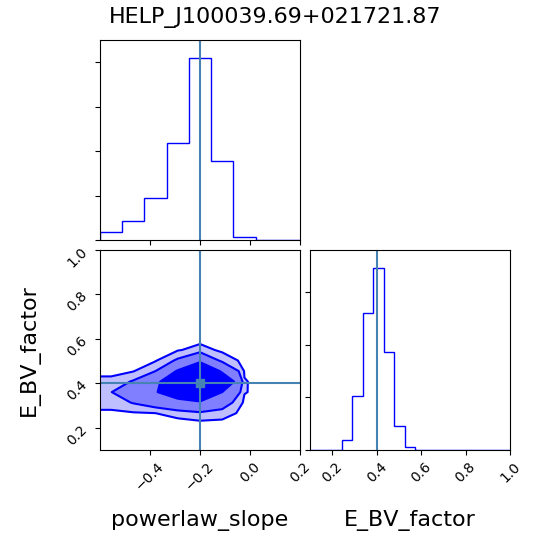}
   \includegraphics[width=0.3\hsize, angle=0]{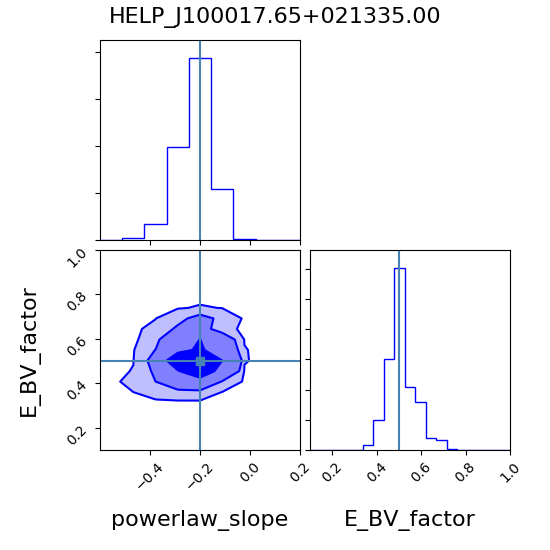}
   \includegraphics[width=0.3\hsize, angle=0]{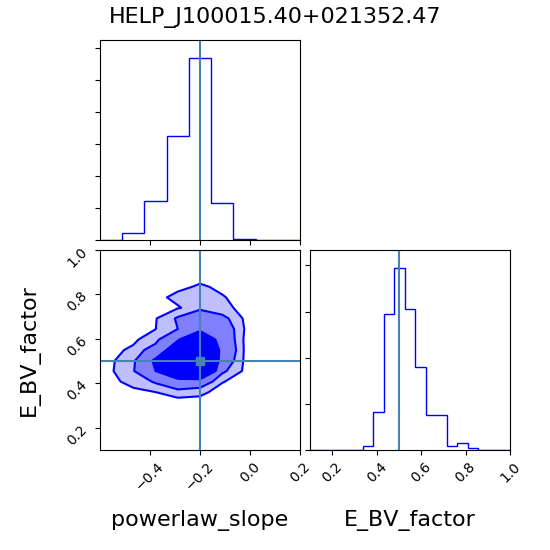}
     \includegraphics[width=0.3\hsize, angle=0]{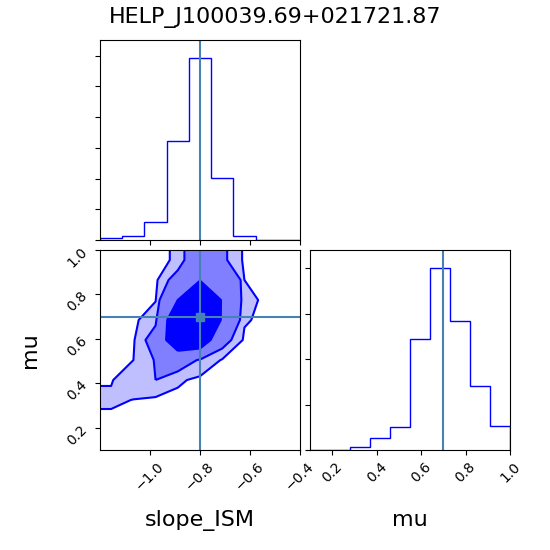}
   \includegraphics[width=0.3\hsize, angle=0]{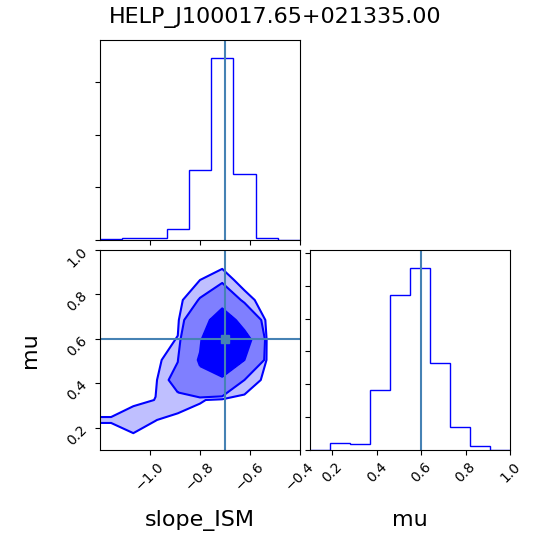}
   \includegraphics[width=0.3\hsize, angle=0]{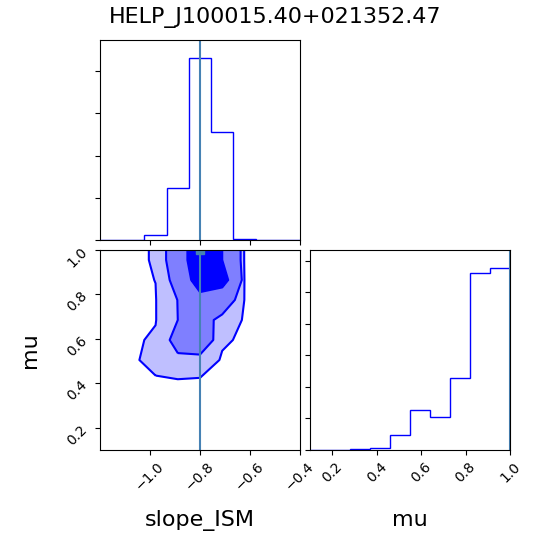}
       \caption{ {Likelihood distributions for three mock datasets between the two intrinsic parameters describing the flexible dust attenuation recipes (see text). The 2D and 1D likelihood distributions are represented in each  panel for the Calzetti-like recipe (upper panels, parameters  $\delta$  (x axis, $\rm powerlaw\_slope$) and $\rm E(B-V)_{star}/E(B-V)_{line}$ (y axis, $\rm E\_BV\_factor$)) and for the DBPL-free recipe (lower panels, parameters $n^{\rm ISM}$ (x axis, $\rm slope\_ISM$) and $\mu$ (y axis, mu)). The contour plots  correspond to 68, 95 and 99$\%$ of the 2D likelihood distributions. The redshifts of the three selected sources are chosen to be representative of the whole sample: from left to right z=0.7, 1 and 1.3. The blue vertical and horizontal lines represent the true input values. The  labels  on the plots are the output parameter names defined in CIGALE.}}
         \label{2Dlike}
   \end{figure*}
\begin{figure}
\includegraphics[width=\columnwidth] {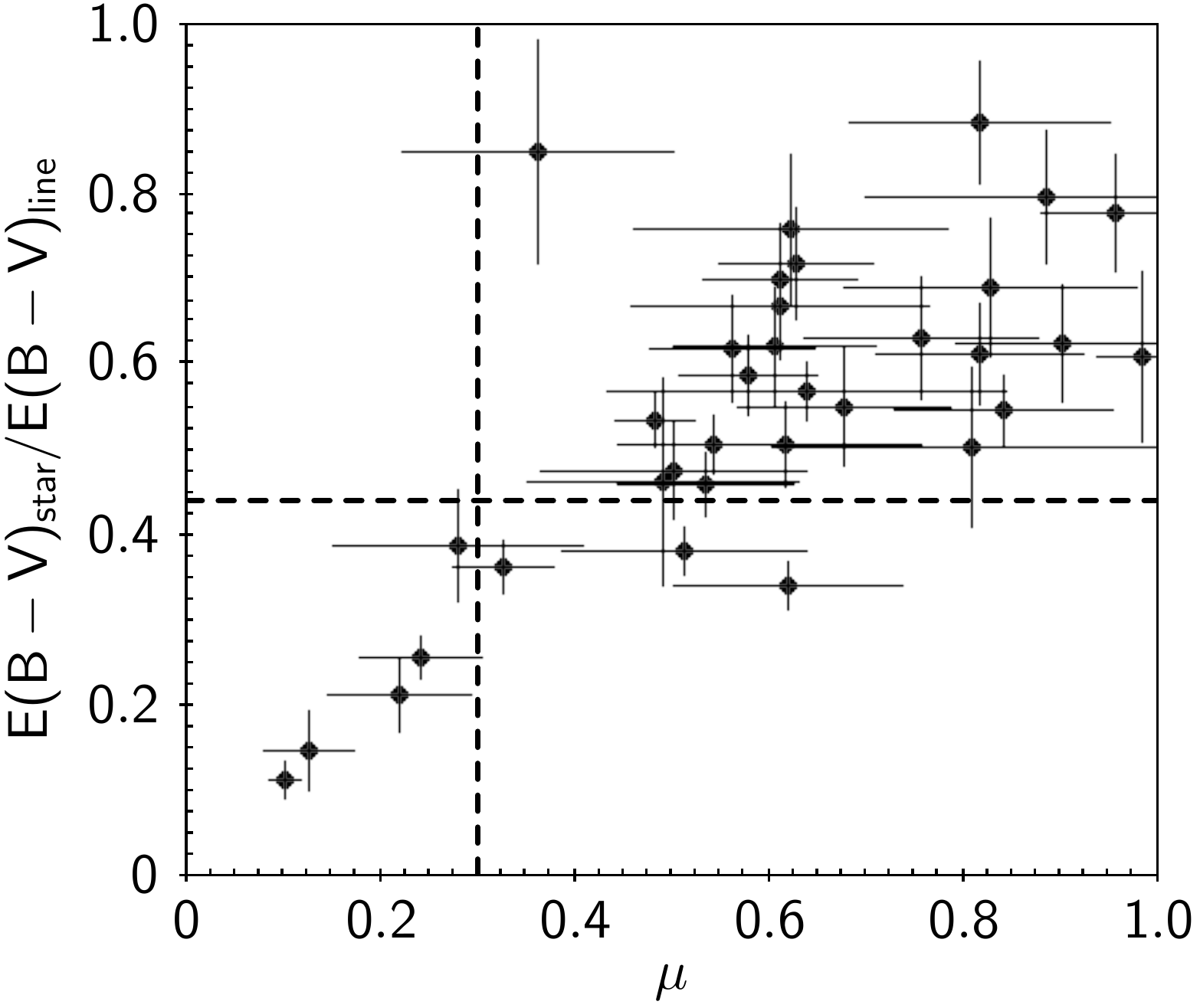}
\includegraphics[width=\columnwidth] {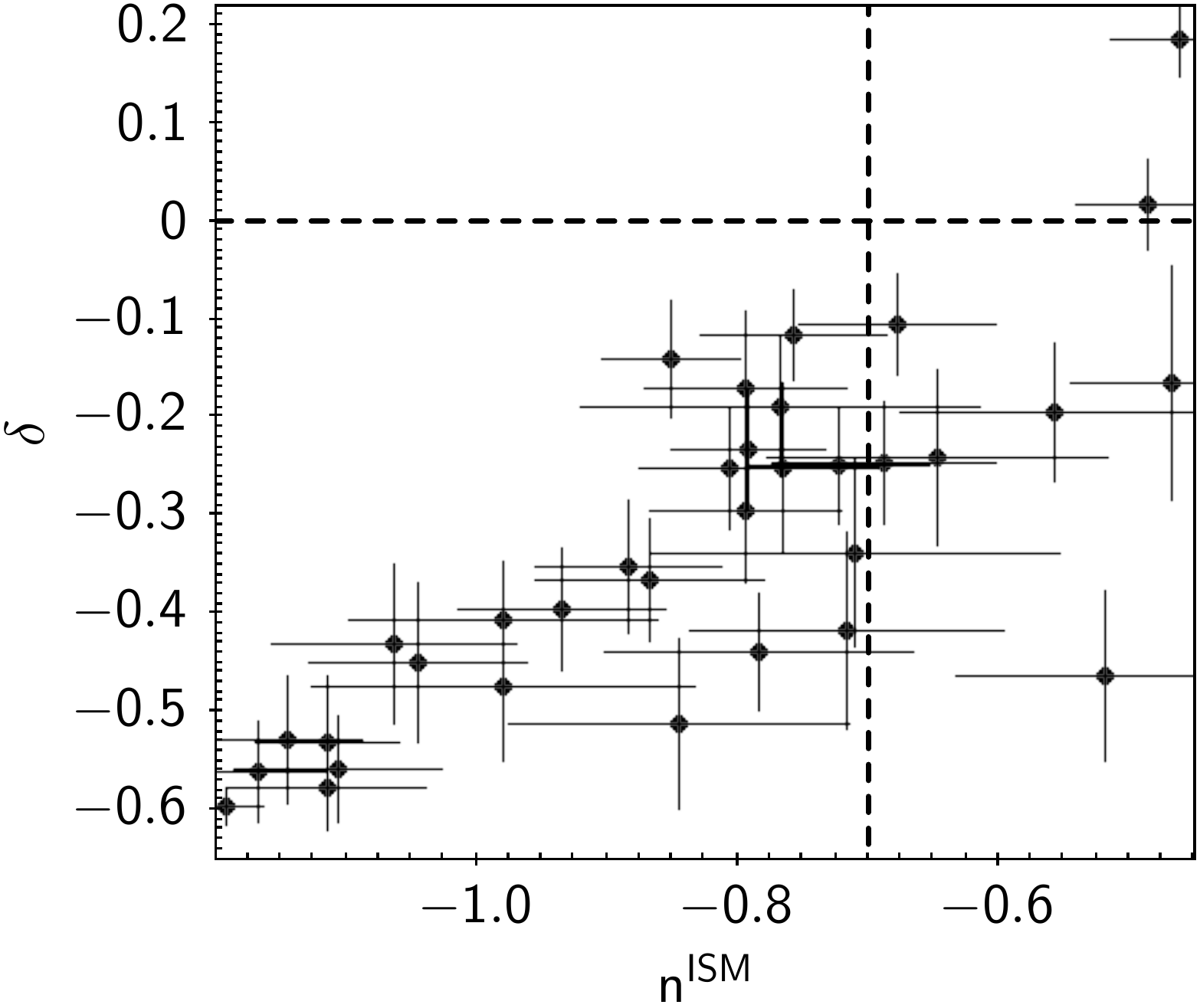}
\caption{Comparison of the  parameters of the attenuation laws. {\it Upper panel:} factor of differential attenuation, $\mu$  (x axis)  against $\rm E(B-V)_{star}/E(B-V)_{line}$ (y axis) for the recipes DBPL-free  and Calzetti-like   respectively.{\it Lower panel:} slopes of the attenuation laws, power-law exponent  $n^{\rm ISM}$  for DBPL-free (x axis) against $\delta$ for Calzetti-like (y axis). In both panels the values of the original recipes C00 and CF00 are indicated with dashed lines.}
	\label{param-attlaw}
\end{figure} 

{The fits performed with the Calzetti-like and DBPL-free recipes introduce two  additional parameters $\delta$ and $n^{\rm ISM}$. Before discussing the  values of these parameters and their relation with other dust attenuation characteristics, it is important to check the accuracy of their measurement and their potential degeneracy with the two other intrinsic free parameters describing the attenuation law, $E(B-V)_{\rm star}/E(B-V)_{\rm line}$ and $\mu$. To this aim we generate mock observations for three objects of our sample at representative redshifts of 0.7, 1 and 1.3 from the best fit of each object. Each flux, calculated by integrating the best fit SED  in the  transmission curve of the filter, is modified by adding a value taken from a Gaussian distribution with the same standard deviation as the observed flux (Boquien et al., submitted). These mock data are  then analyzed in the same way as the real data.  The 2D likelihood distributions  corresponding to the two pairs of intensive parameters ($\delta$, $\rm E(B-V)_{star}/E(B-V)_{line}$) and ($\mu$,  $n^{\rm ISM}$) are presented in Fig. \ref{2Dlike}. with the 1D likelihood distribution of each parameter. No severe degeneracy is found between the parameters.

Another potential issue is the presence of a UV bump in the attenuation curve which may affect the determination of  dust absorption parameters and in particular the slope of the attenuation curve.
It is possible to introduce a UV bump in the Calzetti-like module and such a prescription was already used in various studies \citep{Buat12, Kriek13, Zeimann15}. We perform a  run with a bump of free amplitude. A  positive value of the amplitude is returned  for most objects with an average  value corresponding to $\sim 40\%$ of  the value  found for the MW extinction curve. With at most one broad-band filter overlapping the UV bump, its detection is not safe and we can only say that we do not exclude the presence of a bump. The impact on the measure of $\delta$  when the bump is considered is a very slight average increase of its values  by  0.04$\pm 0.09$ . It never exceeds 0.2 except for two objects for which the difference is of the order of 0.3. For these two SEDs the u band filter overlaps the UV bump and  an amplitude for the bump   as large as the one  of the MW extinction curve is returned by the fit,  while  there is  no valid NUV flux to constrain the slope of the attenuation curve at shorter wavelengths. Such a configuration (positive bump from the fit and no valid NUV data) is  realized only for these two objects. As expected the impact of the presence of the UV bump  on the estimation of  $\rm E(B-V)_{star}/E(B-V)_{line}$  is less important with an average difference between the value obtained with and without the bump  of $-0.02\pm 0.06$. We therefore conclude that the measurement of the dust attenuation parameters introduced in the flexible recipes is  robust.}

Both  parameters $\mu$  and $E(B-V)_{\rm star}/E(B-V)_{\rm line}$ measure  the differential attenuation affecting young and old stars with a slightly different definition as explained in section 3.3.  In the upper panel of Fig. \ref{param-attlaw},  $\mu$ and $E(B-V)_{\rm star}/E(B-V)_{\rm line}$ are both found in general higher than the  values originally adopted  by \citet{Charlot00} and \citet{Calzetti00}  (0.3 and 0.44 respectively). They   have a correlation coefficient of 0.72. Although the two quantities are both related to differential attenuation between young and old stellar populations, we do not expect a 1:1 relation since their definition as well as the separation between young and old stars are different. In the case of Calzetti-like recipe the nebular  and stellar emission are considered separately,  only very young and massive stars contribute efficiently to the nebular emission  reddened with $E(B-V)_{\rm line}$ when  $E(B-V)_{\rm star}$ refers to the attenuation of the continuum emission produce by the    stellar population as a whole. The DBPL-free method splits the stellar population in a young and old component whose attenuation is different and $\mu$ is defined as $A_{\rm V}^{\rm ISM}/ (A_{\rm V}^{\rm ISM}+A_{\rm V}^{\rm BC}$) which is  the ratio of  the  attenuation applied to the old and young stars.   Using similar quantities, $\mu$ can  be compared to $A_{\rm V}^{\rm star}/A_{\rm V}^{\rm line}$\footnote{$A_{\rm V}^{\rm line}$ is obtained by multiplying $E(B-V)_{\rm line}$ by $R_{\rm V}=3.1$ (MW value) and $A_{\rm V}^{\rm star}$ is an output of CIGALE}: these two quantities  are found to strongly correlate ($R=0.83$) with a   mean  ratio of $\mu/( A_{\rm V}^{\rm star}/A_{\rm V}^{\rm line})=1.08\pm 0.19 (1\sigma)$.

 \begin{figure}
\includegraphics[width=\columnwidth] {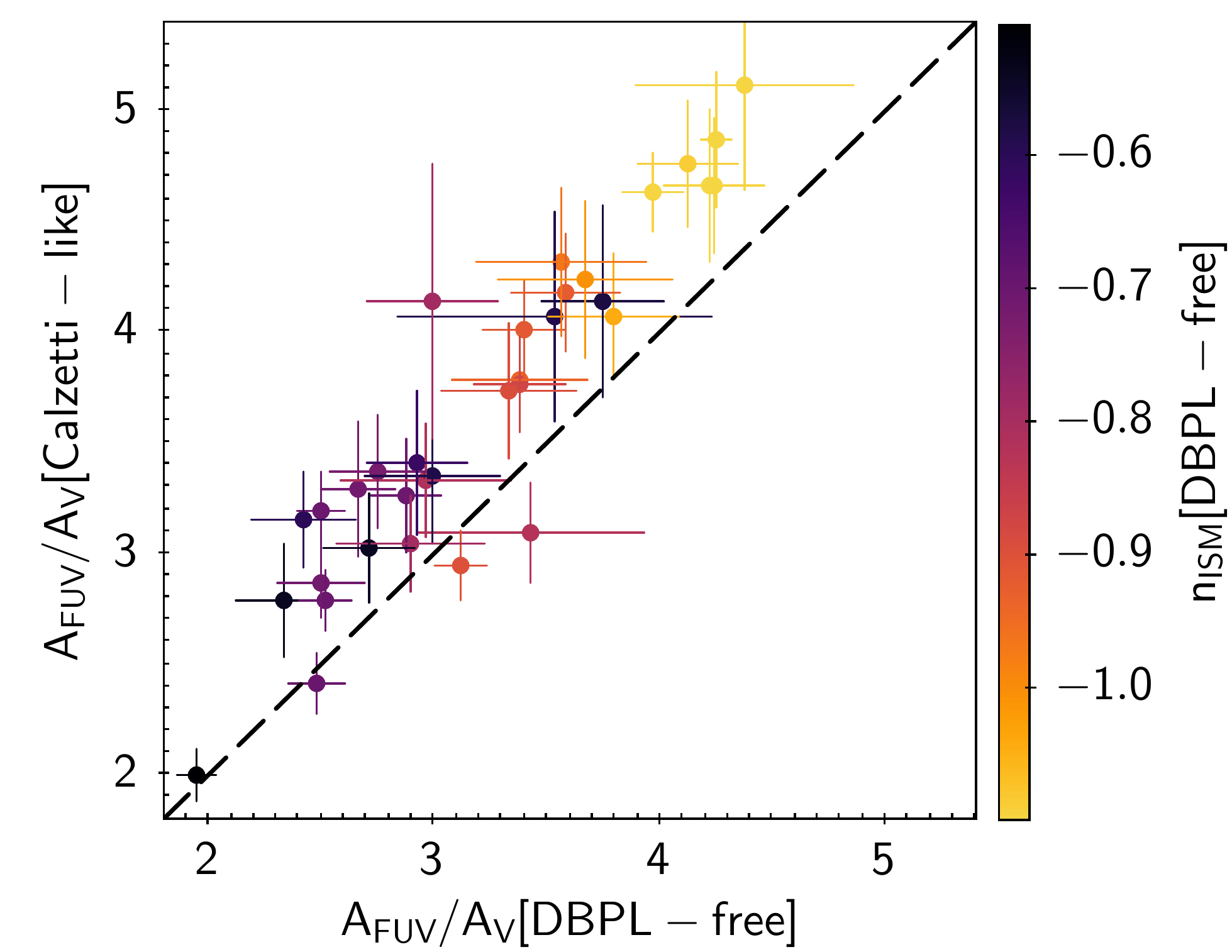}
\includegraphics[width=\columnwidth] {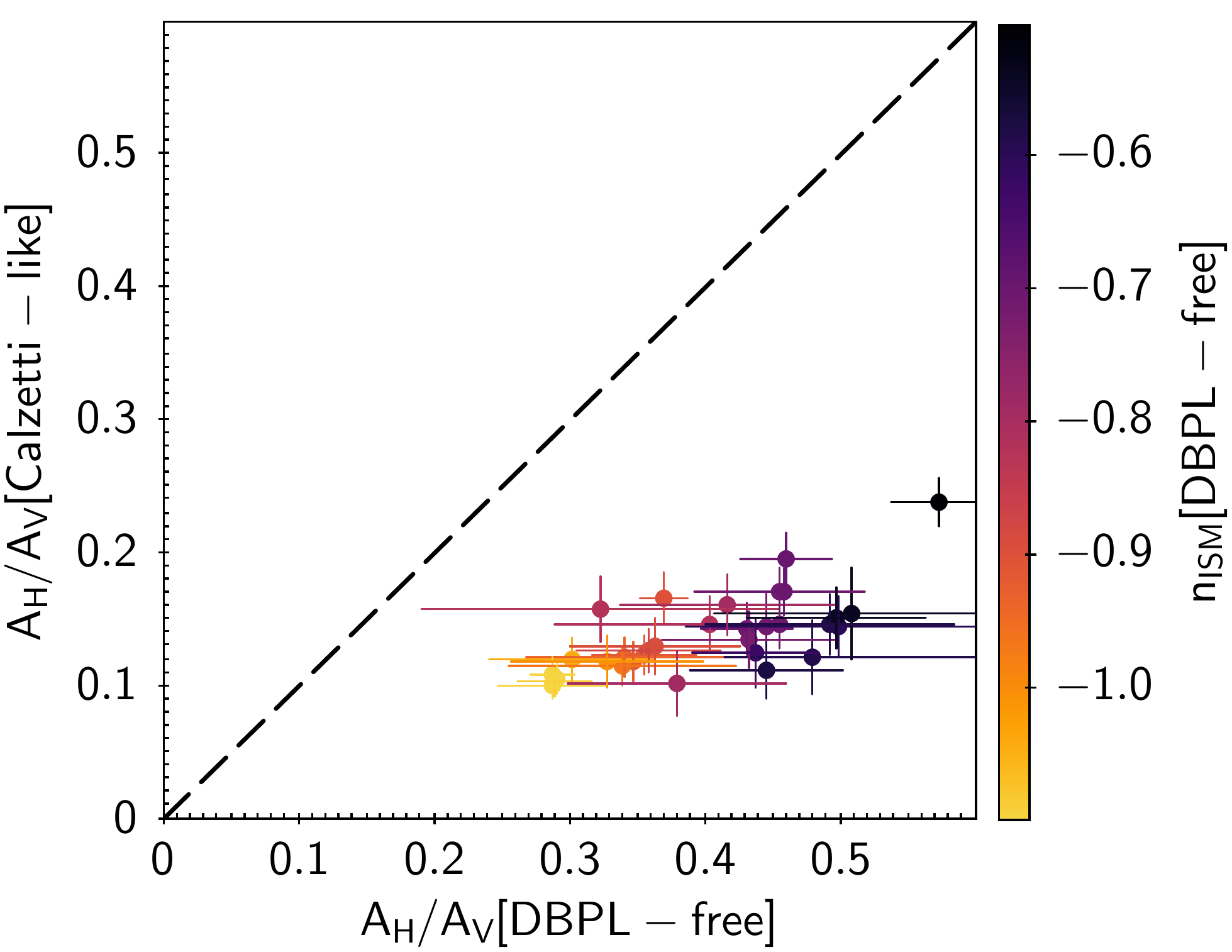}
\caption{ {\it Upper panel}: comparison of values obtained for $A_{\rm FUV}/A_{\rm V}$   with the flexible recipes for dust attenuation DPBL-free (x axis) and Calzetti-like (y axis). The power-law exponent of the attenuation law in the ISM for the DBPL-free recipe is color coded. {\it Lower panel}: same  with $A_{\rm H}/A_{\rm V}$}.
\label{attlaw}
\end{figure}
\begin{figure}
\includegraphics[width=\columnwidth] {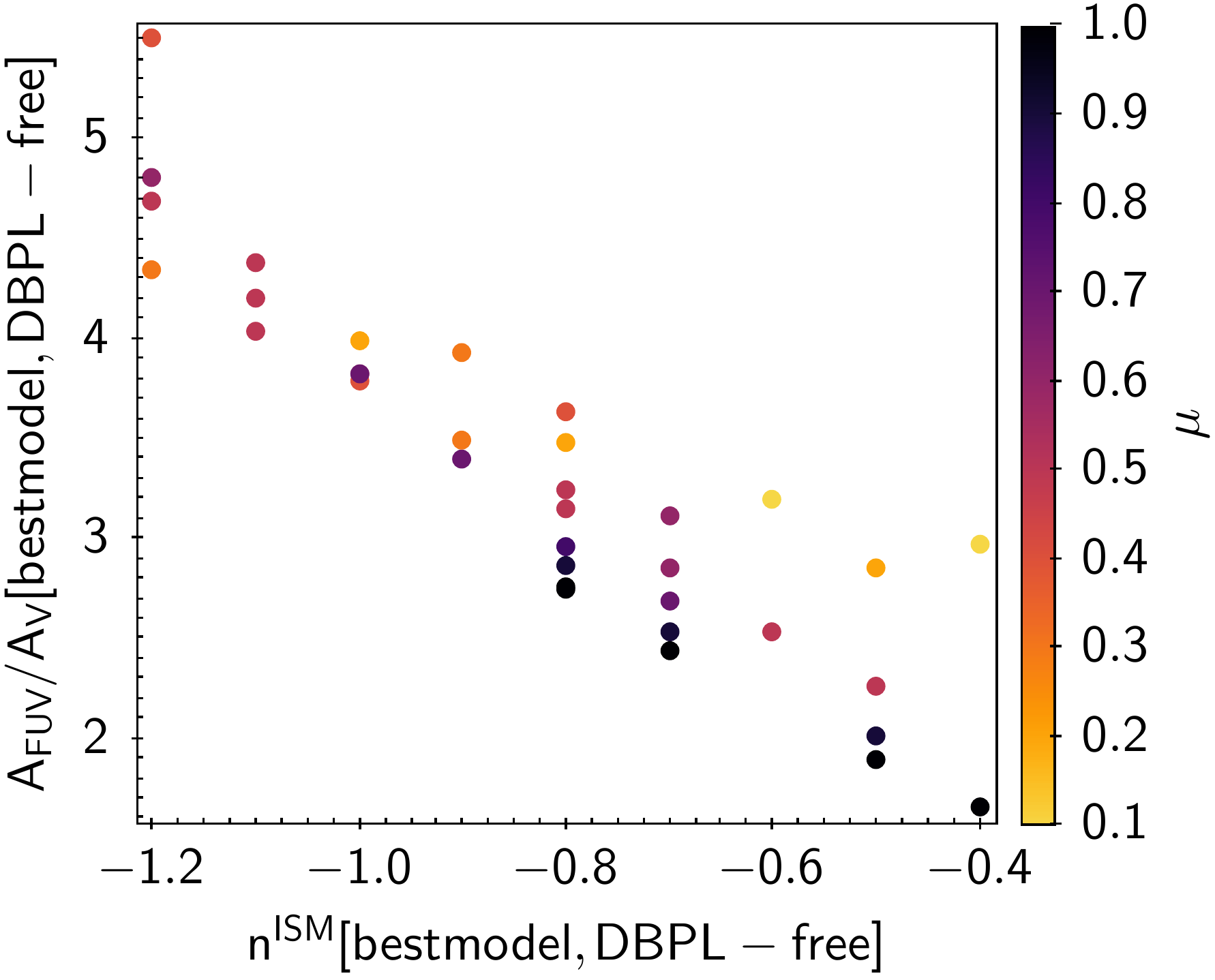}
\caption{ Ratio of the attenuation in the FUV and V bands as a function of    $n^{\rm ISM}$ for the recipe DBPL-free.  The ratio of the V band attenuation experimented by old and young stars,   $\mu$, is color coded. The parameters of the best models are used instead of the means of the PDFs, in order to get clear trends between them}
	\label{dbplfreelaw}
\end{figure}

The exponent $n^{\rm ISM}$ of the DBPL-free recipe is also found to correlate with $\delta$,  which modifies the C00 law, with a correlation coefficient of 0.74 (lower panel of Fig. \ref{param-attlaw}). The  values of $\delta$ are found lower than 0 for all but two galaxies and  only 8  galaxies  are fitted with  an average $n^{\rm ISM} <-0.7$.
While the value of $\delta$ is sufficient to  get  the effective attenuation curve to apply to  the stellar population with our Calzetti-like model, the shape of the effective attenuation  curve resulting from the DBPL-free model depends not only on  $n^{\rm ISM}$ but also on  $\mu$  and on the SFH since  young and old populations are attenuated in different ways. In order to easily  compare the relative variation of the attenuation laws with wavelength obtained with both recipes, we calculate the total attenuation in the FUV, V and H band filters ( with central wavelength  0.15, 0.55 and 1.6 $\mu$m respectively) in order to study  the UV-to-visible and visible-to-NIR    regimes.

In Fig. \ref{attlaw}  the ratios of the attenuation in FUV and V, and H and V respectively are compared.  
It can be seen that the Calzetti-like recipe returns a steeper curve (i.e. a higher $A_{\rm FUV}/A_{\rm V}$ and a lower $A_{\rm H}/A_{\rm V}$). The difference is much  stronger in the visible-to-NIR : $A_{\rm H}/A_{\rm V}$ is found similar  for all the sources fitted with the Calzetti-like recipe: adding a powerlaw dependence does not change the shape of the law in the visible-to-NIR  as  already shown by \citet{LoFaro17}. 
The DBPL-free attenuation laws are  always flatter than the Calzetti-like  laws  in the visible-to-NIR  and, as expected, become greyer ($A_{\rm H}/A_{\rm V}$ increasing) as  $n^{\rm ISM}$ increases.\\
We illustrate the dependence of the DBPL-free attenuation law on $\mu$  by plotting the variation of $A_{\rm FUV}/A_{\rm V}$  with $n^{\rm ISM}$ and $\mu$  for our best models in Fig. \ref{dbplfreelaw}, a clear trend is seen with an increase of  $A_{\rm FUV}/A_{\rm V}$  when $\mu$ decreases, for a given  $n^{\rm ISM}$. We expect a large impact of the differential attenuation on the effective attenuation curve in the UV-to-visible range \citep[e.g.,][]{Inoue05}. When $\mu = 1$, all the stars are obscured with the same  attenuation law which  is a single power-law of exponent $n^{\rm ISM}$. As  $\mu$ decreases, the young stars emitting at lower wavelengths than the older ones are more attenuated: the attenuation law becomes steeper  and $A_{\rm FUV}/A_{\rm V}$ increases. The SFH  also plays a role as shown in \citet{Charlot00}, in the present case our galaxies experiment similar  star formation histories (no strong burst, delayed star formation rate). Nevertheless,  an  effect of the recent star formation activity  can be seen in Fig. \ref{dbplfreelaw} where  $A_{\rm FUV}/A_{\rm V}$ does not have a strictly monotonic variation   with $\mu$.

\subsection{Recovering the H$\alpha$ emission}
\begin{figure}
\includegraphics[width=\columnwidth] {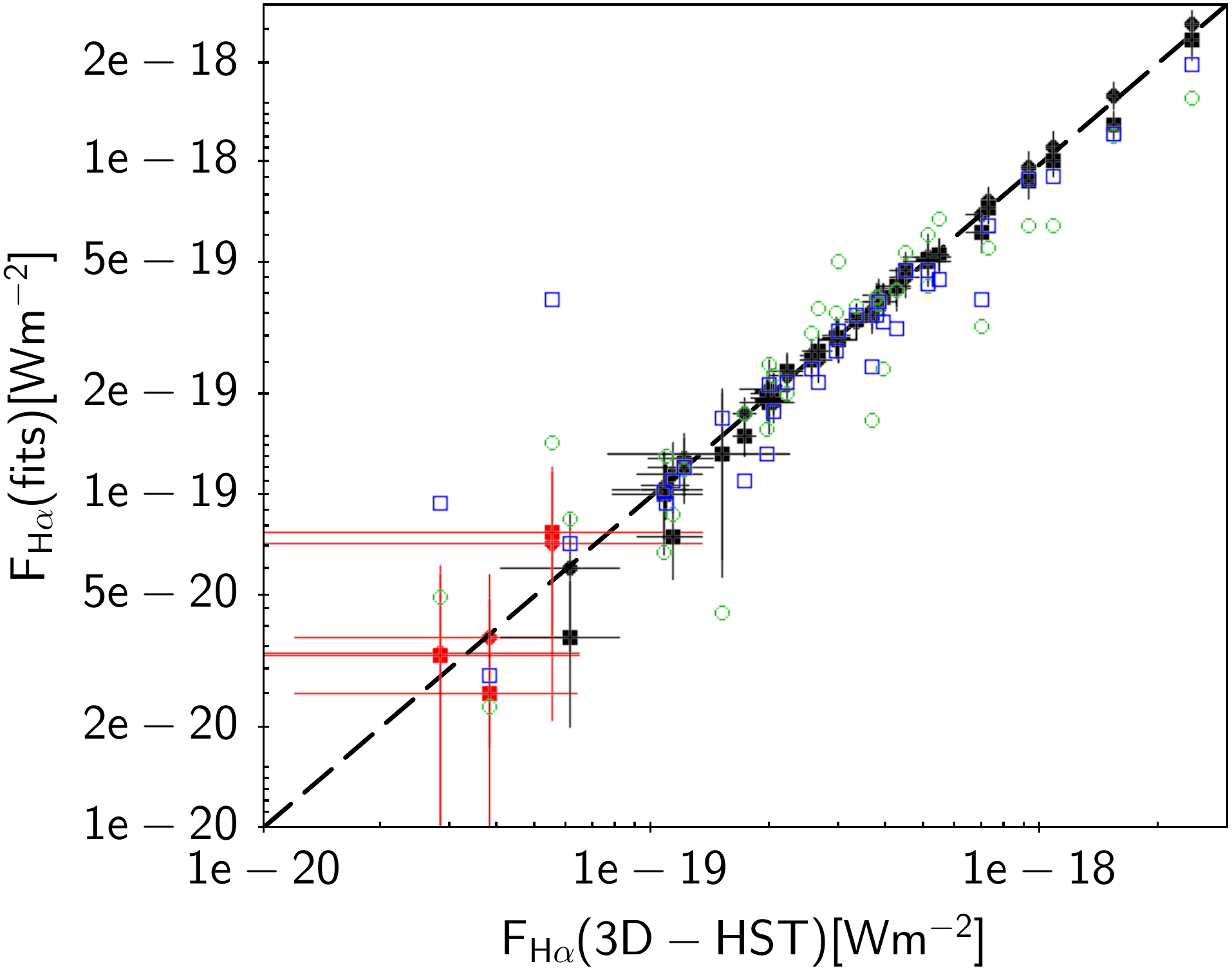}
\caption{{Comparison of  observed 3D-HST H$\alpha$ fluxes} (x axis) with H$\alpha$ fluxes estimated from the fits (y axis). The 4 attenuation recipes are considered: DBPL-free (filled squares), Calzetti-like (filled circles), CF00 (blue empty squares),C00 (green empty circles). The sources detected in  $H\alpha$ with a signal to noise ratio lower than 2 are represented in red with the same symbols as mentioned before. The dashed line is the 1:1 relation. The 1$\sigma$ uncertainty on both measurements (observation and fit) is reported for only two  flexible recipes  (DPBL-free and Calzetti-like)  to make the plot more clear.}
	\label{HAflux}
\end{figure}

{In Fig. \ref{HAflux} the  observed 3D-HST H$\alpha$ fluxes  are plotted against the flux  deduced from the fits  for the four attenuation scenarios}. A very good agreement is found for the sources detected in H$\alpha$ with a signal to noise ratio higher than 2. As expected the dispersion of the correlation between the observed and estimated fluxes is lower with flexible attenuation recipes. The  three galaxies  marginally detected in  H$\alpha$ with a signal to noise ratio lower than 2 are  represented in red in Fig. \ref{HAflux}. Accounting for the uncertainty of the observed H$\alpha$ flux, the observed and fitted fluxes are consistent within 1$\sigma$.

\begin{figure}
\includegraphics[width=\columnwidth] {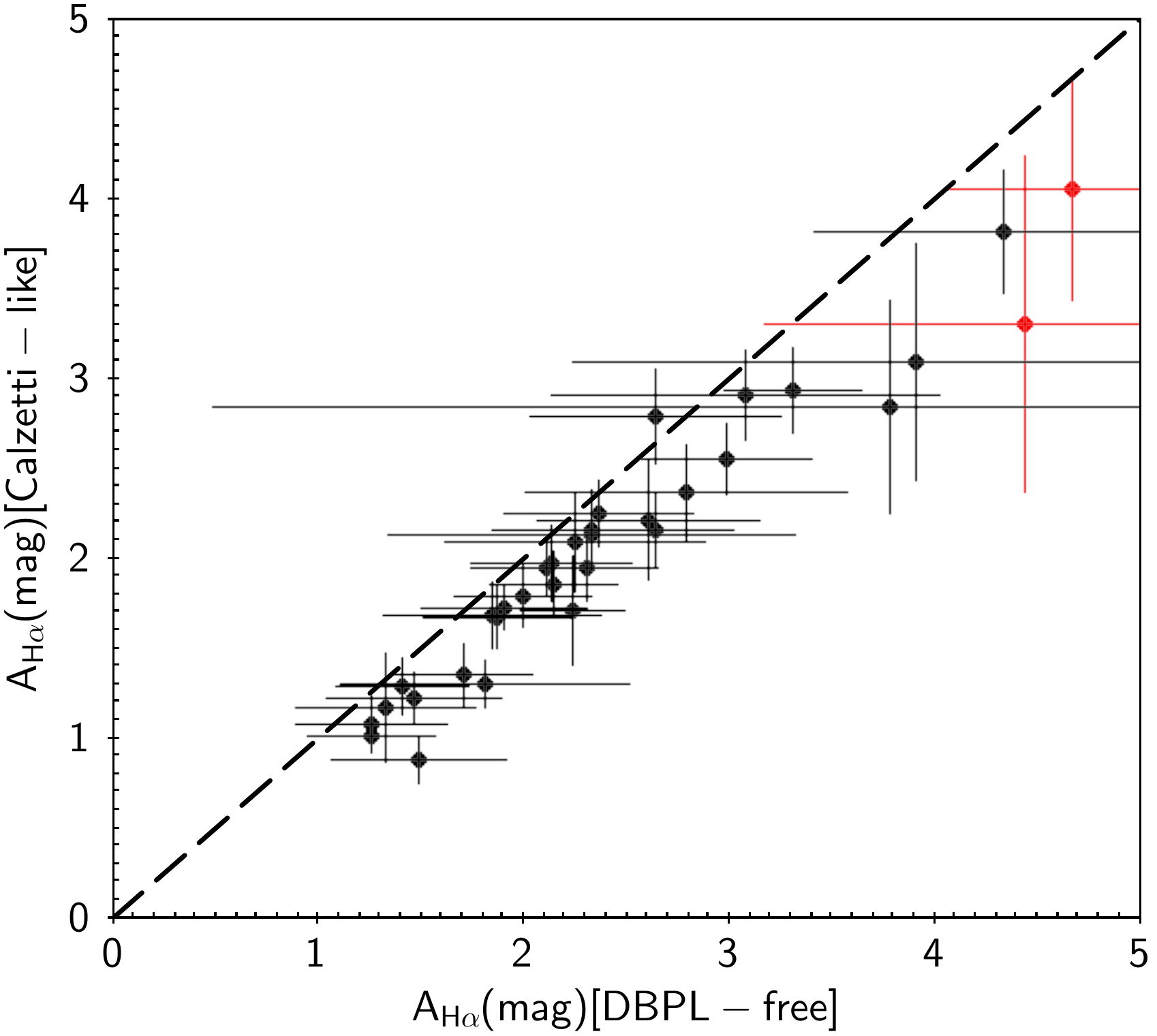}
\caption{ Comparison of H$\alpha$ attenuations obtained with the recipes DBPL-free (x axis) and Calzetti-like (y axis). The 1$\sigma$ uncertainty on both measurements  is reported. The two  sources  represented in red are detected in  $H\alpha$ with an SNR<2, a third source also detected with an SNR<2 is not represented (see text).}
	\label{HAatt}
\end{figure} 
The attenuations in H$\alpha$ obtained with the two scenarios are compared in Fig. \ref{HAatt}, a good agreement is found for the galaxies detected in H$\alpha$ with a signal to noise ratio larger than 2.  The average value of $A_{\rm H\alpha}$, is found to be $2.35\pm 0.75$ and $2.00\pm0.68$ mag for the DBPL-free and Calzetti-like recipes respectively. It is  around twice the usual value found for nearby H$\alpha$ selected galaxies  \citep{Garn10}.  One of the three sources  with a SNR<2 in H$\alpha$ is not represented in Fig. \ref{HAatt}: it exhibits very discrepant $A_{\rm H\alpha}$,   the DBPL-free modeling gives a H$\alpha$ attenuation of $9.2\pm5.7$ mag against  $3.7\pm0.1$ mag with the Calzetti-like model. Despite this large discrepancy, the low impact of the H$\alpha$ flux in the fits (due to its large error) as well as the absence of very young stars (no recent burst) makes the other measurements like the SFR or the attenuation in the FUV or the V band consistent.

\section{Discussion}

\begin{table}
\scriptsize
\centering
\caption{Average characteristics of the attenuation laws with the recipes DBPL-free and Calzetti-like.}
\begin{tabular}{c c c}
\hline
recipe & parameter & mean value \\
\hline
DBPL-free & $n^{\rm ISM}$ & $-0.82\pm 0.19$ \\
DBPL-free & $\mu$ & $0.60\pm 0.23$\\
DBPL-free& $A_{\rm V}$ &  $1.46\pm 0.52$ mag\\
DBPL-free& $A_{\rm H\alpha}$&$2.35\pm 0.75$ mg\\
\hline
Calzetti-like & $\delta$&$-0.33 \pm 0.18$\\
Calzetti-like & $E(B-V)_{\rm star}/E(B-V)_{\rm line}$& $0.54\pm 0.19$\\
Calzetti-like & $E(B-V)_{\rm star}$ & $0.42\pm 0.14$ mag\\
Calzetti-like & $A_{\rm V}$ &  $1.36\pm 0.48$ mag\\
Calzetti-like & $A_{\rm H\alpha}$&$2.00\pm 0.68$ mag\\
\hline
\label{averagelaws}
\end{tabular}
\end{table}
\subsection{Average attenuation properties}
Our fitting method with flexible attenuation recipes allows us to estimate simultaneously  the shape of the attenuation laws and the amount of attenuation for each emission component defined in the recipe adopted for the fit. Moreover, the IR data strongly constrain the measure of the attenuation  through the energy balance. Previous studies measured either the differential attenuation \citep{Kashino13,Price14,Puglisi16}  or only fit  the shape of the attenuation law \citep{Salmon16,Salim18}. The analyses based on spectroscopy, both  for the continuum and the emission lines,  are able to measure all the characteristics of the attenuation process, with methods similar to that of  \citet{Calzetti94} \citep{ Reddy15,Battisti16}. However a robust  absolute calibration of the law should  be based  on the IR emission from dust reemission as performed  by \citet{Calzetti00}. 

The average values of slope, differential attenuation, and absolute values of V band  and H$\alpha$ attenuations found for our sample of galaxies are given in Table \ref{averagelaws}. The average  slope   corresponding to  the Calzetti-like recipe is close  to the average value found by \citet{Buat12}  for a sample of galaxies selected in UV and observed in IR, and steeper than the original law of C00. DBPL-free  gives  slighter flatter slopes than Calzetti-like modeling (cf. Fig. \ref{attlaw}) and  the average value is consistent within 1$\sigma$ with the original value of CF00. The difference between the average slopes of our two flexible recipes  is explained by the introduction of substantial visible-to-NIR  attenuation with DBPL-free which leads to less attenuation in the FUV. The consistency with the CF00 power-law exponent explains why the fits are only slightly improved when $n^{\rm ISM}$ is taken free (section 4 and Fig. \ref{chi2}), it also justifies to use the CF00 value of -0.7 for the entire HELP datasets.

\subsection{Comparison with previous measurements}
 The comparison of our results to the ones already published is possible with the Calzetti-like modeling adopted in several studies, where most of the authors also compare their results to the original C00 law.   \citet{Kriek13}  built composite SEDs including H$\alpha$ features and derived UV slopes ($\delta$ parameter) and UV bump from their fits. They found that $\delta$ is increasing with the H$\alpha$ equivalent width. Our sample spans a moderate range of H$\alpha$ equivalent widths, from 10 to 40 \AA,   with a median value of 13 \AA$~$and our average value of $\delta$ is fully consistent with their measurements. {At higher redshift, \citet{Zeimann15} also  used the Calzetti-like formalism, they  found no evidence of a UV bump and shallow attenuation laws ($\delta>0.2$) for galaxies with stellar masses lower than $\sim 10^{10}$ M$_{\sun}$.  \citet{Salmon16} also found that most of their galaxies have steeper slopes than the C00 law.  However  the average relation they found between the color excess and $\delta$ (their Fig. 9) leads to a	value of $\delta$ close to 0 for  our average attenuation ($A_{\rm V}=1.36\pm 0.48$ mag with the Calzetti-like recipe).  Slopes as steep as ours  correspond to  lower color excess ($\sim 0.1$ mag)}.  {\citet{LoFaro17} used the DBPL-free recipes to fit ULIRGs and found very flat attenuation curves for  these galaxies which experience  a strong  attenuation with an average $A_{\rm V}$ of 2.5 mag, in agreement with a flattening of the attenuation curve when the attenuation increases \citep{Salmon16}.}
 
The average ratio of the color excess for the stellar and nebular emission is consistent within 1$\sigma$  with the 0.44 value valid of nearby starbursting galaxies.  The ratio, $\mu$, of the V-band attenuations for old and young stars has an average value twice the original value proposed by CF00. Several studies  at different redshifts, all assuming a C00 law also found values consistent with  or close to 0.44 \citep{Garn10-2, Price14, Mancini11, Wuyts11} whereas other ones \citep{Kashino13,Puglisi16} found more similar attenuations for nebular and stellar emission. The comparison between these studies is difficult since most of them apply the C00 law to both emissions. If absolute attenuations are measured from comparisons of  SFRs \citep{Garn10-2, Mancini11, Wuyts11} the original value of C00 corresponding $E(B-V)_{\rm star}/E(B-V)_{\rm line}$  must be  translated from 0.44 to 0.57 \citep[e.g.,][]{Pannella15}. The direct measurements of color excesses with Balmer decrement and SED fitting are not affected by this issue \citep{Kashino13, Price14}. The sample used by  \citet{Puglisi16} is also built from 3D-HST and {\it Herschel} data and could be considered close to our selection although they  have PACS  data for  only half of their sample. They found that a value of 0.93, higher than 0.57, has to be used to make consistent the SFR deduced from the UV, IR, H$\alpha$  data. In addition to a different sample selection, the   analyses also differ  with a comparison of SFR based on standard recipes versus SED fitting with a large number of parameters.  \citet{Puglisi16} use the C00 law for their comparison when we also fit the shape of the attenuation curve. The variation of the attenuation curve is not considered in any of the works mentioned above. When we compare the values of $E(B-V)_{\rm star}/E(B-V)_{\rm line}$  obtained with our C00 and  Calzetti-like recipes, the average values are found to be very similar (0.53 against 0.54) but  the difference between the two values of $E(B-V)_{\rm star}/E(B-V)_{\rm line}$ can reach 0.2 on individual objects.
 
Our sample is not well suited to search for correlations between galaxy properties and the parameters of the attenuation law. Its small size and the tight distributions of L$_{\rm IR}$, M$_{\rm star}$,  SFR or sSFR do not give a sufficient statistics to investigate robust trends. \citet{vanderWel12}  measured structural parameters  for  galaxies of the CANDELS survey, the Sersic index of our galaxies spans a small range of values from $\sim 0.5$ to $\sim 2$ and we did not find any significant trend involving this parameter. The axis ratio is better distributed from 0.2 to 0.9 and as expected the attenuation decreases from edge-on to face-on galaxies, although the relation is very dispersed. Anyway, no correlation is found between this axis ratio and the parameters of the attenuation laws.

\subsection{Comparison with  radiative transfer modeling results}
\begin{figure}
\centering
\includegraphics[width=\columnwidth] {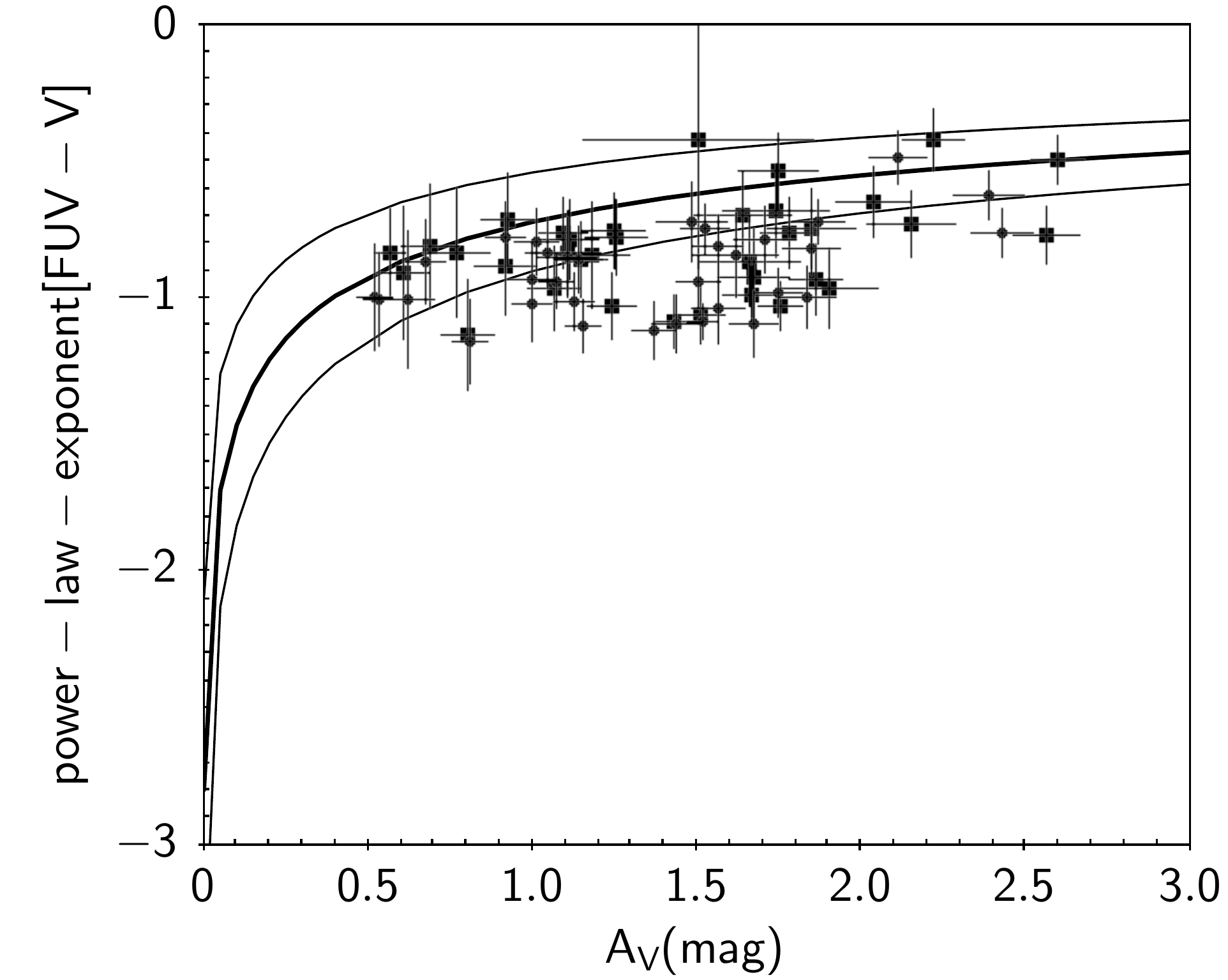}
\includegraphics[width=\columnwidth] {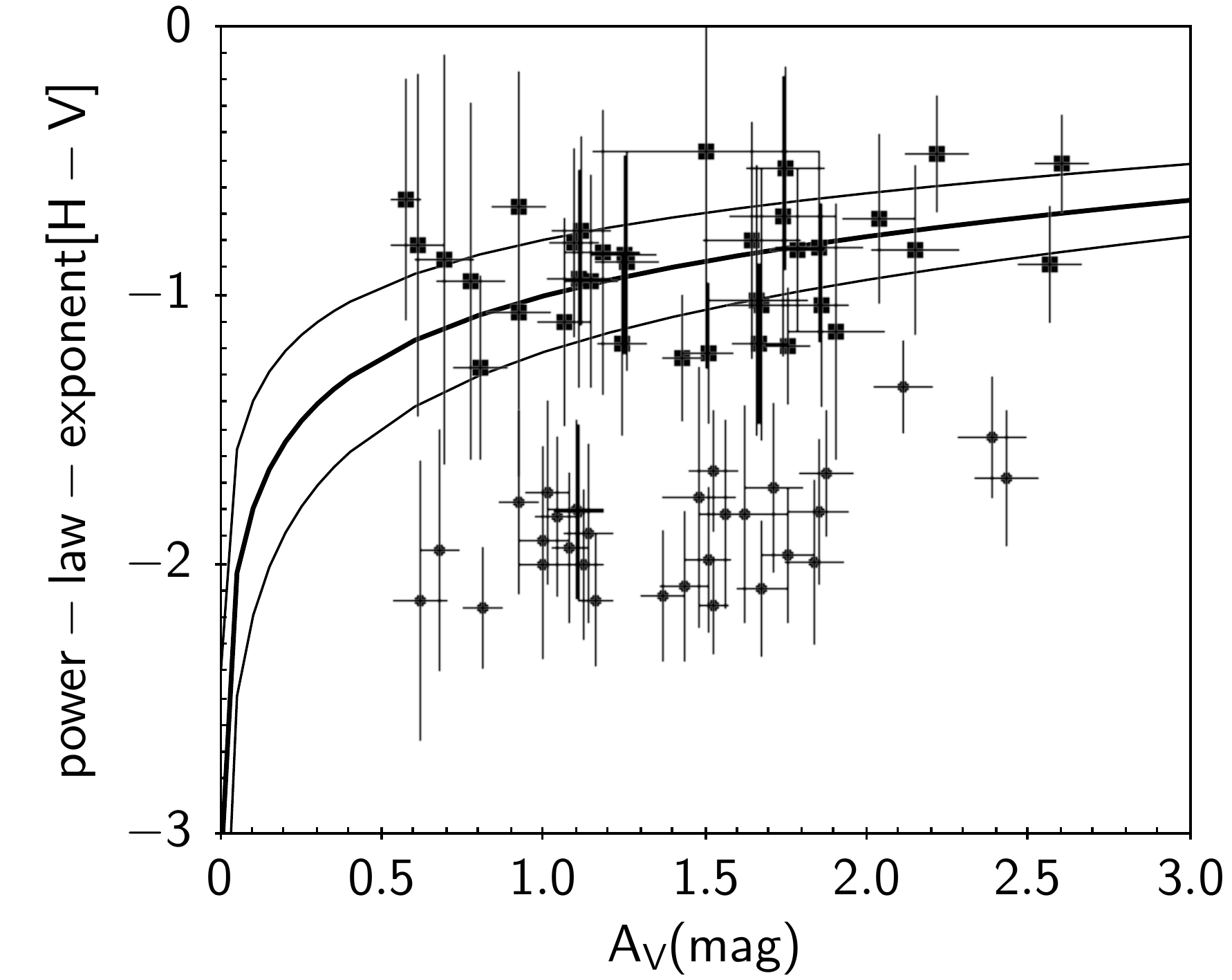}
\caption{The power-law exponents of the effective attenuation laws measured in the visible (upper panel) and NIR (lower panel)  are plotted against the attenuation in the V band, A$_{\rm V}$. and compared to the mean relations of \citet{Chevallard13} plotted with a thick line, the thin lines corresponding to the 1$\sigma$ uncertainty. 
The values obtained with the DBPL-free and the Calzetti-like modeling are plotted with filled squares and dots respectively. The one $\sigma$ uncertainties are reported for each quantity. }
\label{slopes}
\end{figure}

In our analysis  we cannot  distinguish between the Calzetti-like and  DBPL-free modeling, the two scenarios giving fits of similar quality. The consistency of  the  recipes for the estimation of the shape of the attenuation curve at short wavelengths and the good correlation found between the  values of $\delta$ and $n^{\rm ISM}$ on one side and $\mu$ and $E(B-V)_{\rm star}/E(B-V)_{\rm line}$ on the other side   make us confident in  the average trends we find. The two recipes return significant differences in the visible-to-NIR  range as already discussed in \citet{LoFaro17} who found   the DBPL-free recipe to give results more consistent with radiative transfer modeling for dusty ULIRGs. We can also  compare our fitting results  to the predictions of radiative transfer models. \citet{Chevallard13} have gathered the results of several radiative transfer models  to extract the average predicted slope of the attenuation curve, assumed to be a power-law at $\lambda= 0.55$ ($n_{\rm V}$) and 1.6 $\mu$m ($n_{\rm H}$) as a function of the effective attenuation in the corresponding band. To compare with the results of \citet{Chevallard13} we calculate the slope of our  attenuation curves  between the FUV band and the V band, then between the V band and the H band. The choice of such a large difference in wavelength to calculate the power law exponents is justified to minimize the uncertainty on their determination and because of  the global shapes of the attenuation curves obtained with our recipes.   \citet{LoFaro17}  showed that if the DBPL-free recipe leads to attenuation curves close to a perfect power law it is not the case with the Calzetti-like model and the variation between FUV and V on one side and V and H on the other side give a good global representation of  both laws. In Fig. \ref{slopes}, the prediction of the models (Eq. 9 and 10 of \citet{Chevallard13}) and our resulting slopes are compared. A reasonable  agreement is found in the V band for both recipes, our slopes are slighly steeper, especially with the Calzetti-like models. The situation is very different in the NIR: the slopes obtained with the DBPL-free models are in good agreement with the radiative transfer models but the slopes  derived  with the Calzetti-like recipe are much steeper. The introduction of the $\delta$ parameter modifies the slope of the attenuation law in the UV range but  has no significant impact on the shape of the attenuation curve in the NIR range  while  flatter curves  are  expected from radiative transfer models for the effective attenuation found for our sample galaxies.

\section{Conclusion}
We have analyzed a complete sample of 34 galaxies at intermediate redshift selected in IR with {\it Herschel} PACS and SPIRE detections, with an H$\alpha$ line detection for each source from the 3D-HST survey. By fitting simultaneously  UV-to-NIR  broad band  and H$\alpha$ fluxes we have compared several  dust attenuation recipes: the classical \citet{Calzetti00} and \citet{Charlot00} methods  and flexible recipes modifying the slope of the effective attenuation curves. The differential attenuation between either stellar and nebular emission or young and old stars was taken as a free parameter for each recipe .

The flexible recipes give the best results with a substantial variation of the slope of the laws and of the differential attenuation, confirming the non-universality of  any dust attenuation law. The \citet{Charlot00} give satisfactory results in its original form while  the slope of the \citet{Calzetti00} law is substantially  modified. The H$\alpha$ fluxes are very well fitted together with the continuum emission, the mean attenuation in the H$\alpha$ line is found to be of the order of 2 mag. 

The effective attenuation curves obtained with the flexible recipes are found  in general steeper than the original ones. Both recipes are consistent with the general trends found with radiative transfer modeling in  the visible between the  global attenuation and the slope of the effective attenuation  law. Large departures are found in the NIR: the flexible recipe based on \citet{Charlot00} prescription remains consistent with the results of the models but the multiplication of the  \citet{Calzetti00} law  by a power law  does not  sufficiently flatten  the slope of the resulting law in the NIR to reach the values predicted by models.

The attenuation in the V band affecting young stars is found to be, in average, 1.6 times higher than the one found for older stars, a factor 2 lower than the original value proposed by \citet{Charlot00}, the average value of the ratio between the color excess for stellar continuum and the nebular emission is consistent with the value 0.44 valid for the   \citet{Calzetti00} law. However, both parameters are found to span a large range of values.
 
\vspace{3cm}
\paragraph{\textit{Acknowledgement}}
We thank the anonymous referee for her/his very useful  comments on   validity checks to implement. This study has received funding from the European Union Seventh   Framework Programme  FP7/2007-2013/   under Grant Agreement Number 60725. MB was
supported by the FONDECYT regular project 1170618 and part of the work was done during a stay
of VB at Universidad de Antofagasta funded by the FONDECYT project. KM has been also supported by the National Science Centre (grant UMO-2013/09/D/ST9/04030).
This work is based on observations taken by the 3D-HST Treasury Program (GO 12177 and 12328) with the NASA-ESA HST, which is operated by the Association of Universities for Research in Astronomy, Inc., under NASA contract NAS5-26555.

\nocite{*}
\bibliographystyle{aa}
\bibliography{aa-revised1}

\end{document}